 \documentclass[final,5p,times,twocolumn]{elsarticle}

\usepackage{graphicx}
\usepackage{amssymb}
\def\msol{M$_{\odot}$ }
\def\kms{$\rm km\, s^{-1}$}
\def\cm3{$\rm cm^{-3}$}
\def\Ts{$\rm T_{*}$~}
\def\Vs{$\rm V_{s}$~}
\def\n0{$\rm n_{0}$}
\def\B0{$\rm B_{0}$}

\def\erg{$\rm erg\, cm^{-2}\, s^{-1}$}
\def\mum{$\mu$m}

\def\mum{$\mu$m}

\def\L12{L$_{12\mu m}$~}
\def\F12{F$_{12\mu m}$~}

\def\Hb{H$\beta$~}
\def\Ha{H$\alpha$~}

\def\ff{{\it ff}~}

\journal{New Astronomy}

\begin{document}

\begin{frontmatter}



\title{The symbiotic system  AG Draconis\\
Soft X-ray bremsstrahlung from the nebulae}


\author{Marcella Contini$^1$ and Rodolfo Angeloni$^2$}

\address{
$^1$School of Physics and Astronomy, Tel-Aviv University, Tel-Aviv, 69978 Israel\\
$^2$Departamento de Astronom\'{i}a y Astrof\'{i}sica, Pontificia Universidad Cat\'{o}lica de Chile -
Vicu\~{n}a Mackenna 4860,
 7820436 Macul, Santiago, Chile
}

\begin{abstract}
The   modeling of UV and optical spectra  emitted from the symbiotic system AG Draconis,
adopting collision of the winds,  predicts  soft
X-ray bremsstrahlung from nebulae downstream of the reverse shock with velocities $>$ 150 \kms
and  intensities comparable to those of the white dwarf black body flux.
At outbursts, the  envelop of debris, which
 corresponds to the nebula downstream of the high velocity shocks
 (700-1000 \kms)  accompanying the blast wave,
 absorbs the black body soft X-ray flux from the white dwarf,
 explains the broad component of  the H and He lines, and  leads to low optical-UV-X-ray continuum fluxes.
The high optical-UV flux observed at the outbursts is explained by
bremsstrahlung downstream of the reverse shock between the stars.
The depletion of  C, N, O, and Mg relative to H
indicates that they are trapped into dust grains and/or into diatomic  molecules,
 suggesting that the collision of the wind from the
white dwarf with the dusty shells, ejected  from the red giant with  a $\sim$1 year periodicity,  leads to the
U-band fluctuations during the major bursts.

\end{abstract}


\end{frontmatter}

% \linenumbers


\section{Introduction}

Circumstellar and circumbinary nebulae throughout symbiotic systems  (SS)
emit  line spectra and   continuum  in the radio, infrared (IR), optical, UV, 
and X-ray range  (Angeloni et al. 2010 and references therein).

The two main SS components, generally a white dwarf (WD) and a red giant (RG)  rotating in bound orbits,
lead to  characteristic time  periodicities. Moreover, they dominate   the continuum spectral energy distribution
(SED) in the UV-soft X-rays and IR-optical ranges, respectively.

The RG flux  is well reproduced  by a black body (bb).
The  analysis  of symbiotic systems  
 shows that the
 flux from the hot component star is approximated by a bb which peaks in the  far UV.
It  was rarely observed, beyond a few  data in the bb  low frequency tail.
Consequently, the temperature of the hot star  was obtained  indirectly  from the
characteristic spectral line ratios emitted from the nebulae (Angeloni et al. 2010).

AG Draconis (AG Dra) is classified as a yellow symbiotic binary  containing a K2III giant 
(M\"{u}rset \& Schmid 1999)
and a WD accreting from the giant's wind on a 549 day orbit (Fekel et al. 2000).
The orbital inclination is relatively low with  $i\sim$30$^o$-45$^o$ (Mikolajewska et al. 1995) or
$i \sim$ 60$\pm$ 8.2$^o$ (Schmid \& Schild 1997).
In fact, during the orbital motion the  WD is not eclipsed.

The light curve of AG Dra shows numerous bursts with amplitude  spanning 1-3 mag in U.
Gonz\'{a}lez-Riestra et al. (1999, hereafter GVIG99) 
identified "cool" and "hot" outbursts differing in Zanstra temperature and 
light curve profile (Skopal et al. 2009).
Tomov \& Tomova (2002 and references therein) report   from the analysis of the  photometric data  
 that the 
visual light growth during the  optical brightening is due to increased radiation of the
circumbinary nebula. Thermonuclear events  on the surface of the WD (Mikolajewska et al
al. 1995) and expansion and cooling of the compact secondary (Greiner et al. 1997)
were also  conjectured  to explain  the outbursts.

Besides the  characteristic outbursts in the U band,
AG Dra is the brightest system at soft X-ray energies  of all SS.
Viotti et al. (1998) monitored  the super soft X-rays from the 
AG Dra source by EXOSAT and ROSAT and  the UV flux by IUE.
It was found that during all the monitored outbursts the visual magnitude was associated
with an increase in the UV flux, while the X-ray count rate largely faded, namely
the  count rate dramatically dropped at the time of the  1985 and 1986  maxima 
(Viotti et al. 1998) and between 1992 and 1996  (Greiner et al. 1997).
Actually, Viotti et al. (1998)   discovered the systematic
optical/UV and X-ray flux anti-correlation, which was  confirmed by Skopal et al. 
(2009), Gonz\'{a}lez-Riestra et al. (2008), etc.  observations.

Analyzing new data by XMM Newton and  FUSE, Skopal et al. (2009) recently 
concluded that the soft  X-ray emission is produced by the WD photosphere. 
They claim that  X-ray and far UV observations   determine the bb temperature unambiguously.
Modeling the   UV and super-soft X-ray  data  in AG Dra, 
they  concluded that  the  observed anti-correlation is caused by the variable wind from the hot star,
explaining that the enhanced hot star wind gives rise to the optical bursts by reprocessing high
energy photons from the Lyman continuum to the optical/UV.

This paper is motivated by the analysis of AG Dra XMM-Newton observations (Jansen et al. 2001)
performed during the 2003 hot burst and during quiescent phases by Skopal et al. (2009), who
 found that the  soft X-ray spectrum   at quiescence  corresponds to  bb radiation.
 We  would like to investigate   by  consistent
modeling of the line and continuum spectra whether the bremsstrahlung downstream of
shocked nebulae  might contribute to the soft X-ray emission at different phases.

Significant 
 information about the physical conditions of the SS components can be obtained
by  line spectrum analysis.
The intensities and profiles  of some spectral lines  emitted from AG Dra at different epochs
were analyzed in detail (e.g. the \Ha and HeII 4713 lines by Tomov \& Tomova 2002, 
the OVI 1032,1035 doublet and HeII 1640 line by Skopal et al. 2009).
The  most comprehensive and complete  sample of spectra   in number of lines  at different close  
phases in the optical and UV ranges was  provided  by GVIG99.
Both the line and  continuum spectra  were explained by  
 photoionization of circumstellar matter, e.g. the wind from the RG, and of circumbinary matter
by the hot star radiation flux.

 Viotti et al. (1989)  noticed, however,  that  {\it in some models X-ray are interpreted as
bremsstrahlung emission from the collision of the winds from both components
(e.g. Kwoks \& Leahy 1984, Girard \& Willson 1987). The steady and intense X-ray emission of
AG Dra during quiescence indicates the presence of an effective heating process.}

Therefore in this paper we   calculate the spectra  using  composite models which account consistently for shocks 
and photoionization.
We will follow the method  adopted by Angeloni et al. (2010)
for other symbiotic systems,
cross-checking   the calculations of the  continuum by those of  the line ratios.
We will focus on the optical-UV-X-ray domain, leaving the radio-IR
range to the next investigation of the AG Dra SS.

The   models presented  by Skopal et al. (2009)  for AG Dra are valid on  a large scale,
namely the continuum emitted from the nebulae was not constrained
by the analysis of  the line spectra. The collision of the winds cannot be neglected
because it leads to high temperatures of the  gas. In this paper
we would like therefore to investigate  whether, adopting  models accounting for both
photoionization and shocks, it is possible to add some
information to the  interpretation of AG Dra spectral features, in particular, and of
SSs, in general.

The SS modeling  and the calculation of the spectra  are explained in Sect. 2.
The spectra are analyzed in Sect. 3.
Results  are presented in Sect. 4.  Discussion and concluding  remarks  follow in Sect. 5.

\section{The models}

Alternating episodes of  accretion and  wind collision were evident
during the evolution of  symbiotic systems, e.g. CH Cyg.
The epochs treated in this work on   AG Dra are close to outbursts. We consider
that the accretion disk was blown up and  had no time to form.
In fact, the observed spectra do not show the typical
double peaked lines.
So we adopt the colliding wind model of Girard \& Willson (1987), which applies
close to the orbital plane of the SS.

\subsection{Wind collision theory}

 Girard \& Willson (1987)  explained that in  a widely separated 
binary consisting of a late giant  and a hot WD, the late star has a 
stellar wind with a mass loss rate of 10$^{-7}$-10$^{-5}$ \msol yr$^{-1}$ and a 
velocity  $\sim$10-30 \kms.
The hot star will accrete hydrogen rich material 
 from  the late red giant wind. Theoretical models by
Paczynski \& Zytkow (1978) show that an  accretion of hydrogen rich material at a
rate of 10$^{-11}$ to 10$^{-10}$ \msol yr$^{-1}$  will   generate periodic hydrogen 
shell flashes, with
each outburst lasting on the order of decades.
The increased flux from such an eruption, particularly in the UV region, and
the accompanying increase in radiation pressure will tend to eject the hot
star's hydrogen atmosphere in the form of a stellar wind with velocities of
1000-3000 \kms\ (Conti 1978),
which  may last  for decades. 
Both the winds  from the WD and the red giant are supersonic, so shocks will form.

Since the end of the 1990, it was suggested by the  analysis of  SS line  spectra 
(Nussbaumer  2000 and references therein) that both  stars lose winds which collide 
in the interbinary region and outward,
creating   a  network of shock-fronts. 
A picture of the colliding wind model adapted to the presence of dust shells
is given by Contini et al. (2009c, fig. 2).

The   scenario  shows the collision
of the winds from both  stars leading to two  shock-fronts which propagate
between the stars: the strong one, dominating the spectrum,  propagates
in reverse towards
the WD, and the weaker one, not always important,  propagates towards the RG.
Moreover, a shock-front  expands out of the system through the circumbinary
medium.

\subsection{Calculation of the spectra}

 Composite models  considering  both photoionization and shocks in a  
consistent way  were adopted (Contini 1997 - Angeloni et al. 2010).

In the past years we carried out  the  analysis of  SSs (WD + RG
+ nebulae  + dusty shells) of both line and continuum spectra
emitted from the nebulae downstream of  shock-fronts and
 of dusty shells ejected by the RG.
In particular, the observed SED was   disentangled into  the different components
by adjusting model calculations  to the data in the different frequency ranges.
The models were constrained by cross checking the line ratios  with 
 the continuum. Comparison with the  $absolute$ observational data
 determines the  radius of the nebulae throughout the SS.

The collision  of the outburst high velocity  ejecta 
with the RG wind and/or with the ISM  surrounding the WD on the side opposite  the RG
(Contini et al. 2009a),
induces  the  high  temperatures  downstream of the shock-fronts
and, consequently, to  X-ray emission (Angeloni et al.  2010, fig. 4), while
 in  the  UV domain  the WD  bb flux dominates.
However, the data in the X-ray range were not always  sufficient to  distinguish  the  origin of the
far-UV - soft X-ray emission. Both the bb tail from the WD and
bremsstrahlung  emitted from nebulae downstream of relatively  fast  shocks
were  proposed (e.g. Contini \& Formiggini 1999, fig.4).

In particular, the nebulae downstream  are photoionized and heated by the flux from the WD and collisionally
by the shock.
The gas entering the shock-front is compressed and heated in the post shock-region.
The flux from the WD and the secondary diffuse radiation flux
are  calculated throughout a large number of slabs downstream (up to 300)
by radiation transfer.
The gas  cools down by free-free, free-bound and line emission.
The temperature drop is calculated
smoothly downstream  throughout the slabs whose geometrical thickness is automatically
calculated by the temperature gradient. The computation
halts  when the the gas is cool enough
in the radiation bound case, or, in the matter bound case, when  the edge opposite to the
shock-front is reached.

We use for the calculation of the spectra the code
SUMA\footnote{http://wise-obs.tau.ac.il/$\sim$marcel/suma/index.htm},
which simulates the physical conditions of an emitting gaseous nebula under the  joint effect of
 photoionization from an external source and shocks, 
and in which line and continuum emission from gas are calculated consistently with
dust reprocessed radiation as well as with grain heating and sputtering.

\section{The  spectra}

In AG Dra, the broad and narrow  components of the line profiles reveal  high  
(700 to 3000 \kms) and low (100-160 \kms) velocities
 at  the  epochs corresponding  to the outbursts.
The contemporary presence of high and low velocity shocks   between the stars
with a velocity gap by a factor  $>$ 10
indicates that the matter is highly fragmented by  instabilities at the fluid interface
(Richtmyer-Meschkov, Kelvin-Helmoltz, Rayleigh-Taylor) created by the outburst.
Alternatively, the high velocities may
 characterize the shocks
accompanying the  blast wave created  by the outburst  in  directions 
opposite to the RG. These shock-fronts 
can be treated by the dynamical evolution described by Chevalier
(1982)  in  supernova  atmospheres (Contini et al. 2009a).
We will see in  the following that the latter hypothesis gives a more  acceptable
explanation of   AG Dra observations.

Following our method,  first we model the line ratios, which  constrain the model,
then we cross check   the   results  by comparing the calculated continuum SED  with the data.  

To constrain the models, the spectra  should contain enough   significant lines, such as
those from the same element but  from different ionization levels  which give information
about the conditions of the emitting gas independently of relative abundances,
and lines from the same level of different elements  which lead to different
stratifications of the ions downstream, depending on the ionization potential.
Comparison of line ratios in the UV and in the optical range also helps
 understanding  the gas conditions within the  SS.

\begin{figure*}
\begin{center}
\includegraphics[width=0.70\textwidth]{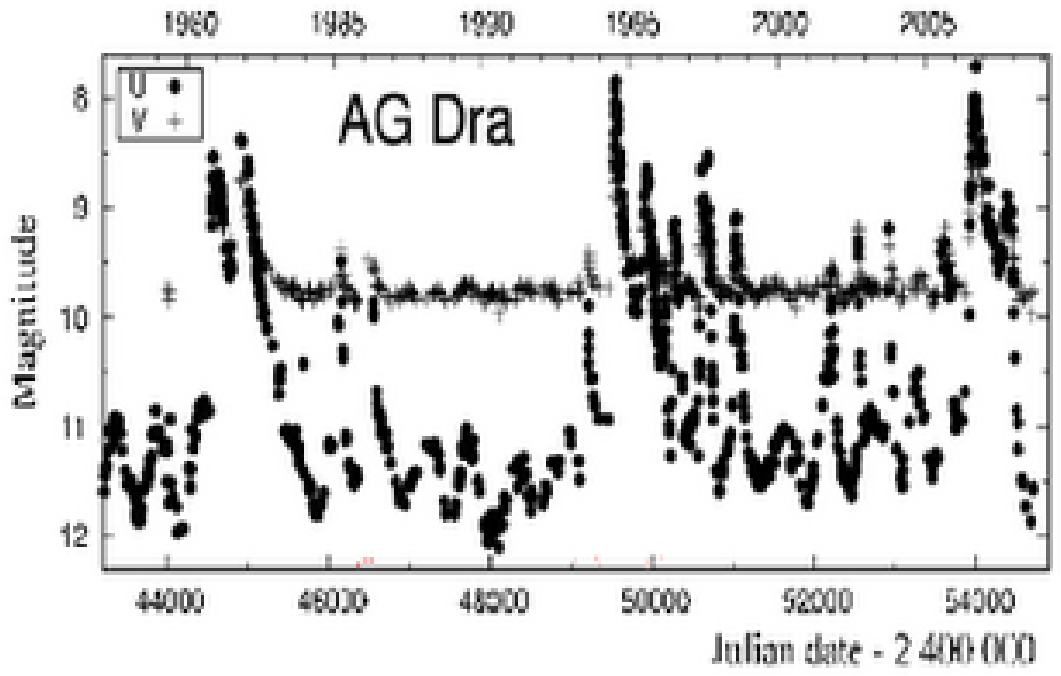}
%\includegraphics[width=0.50\textwidth]{Fig1b.eps}
%\flushleft
\includegraphics[width=0.35\textwidth]{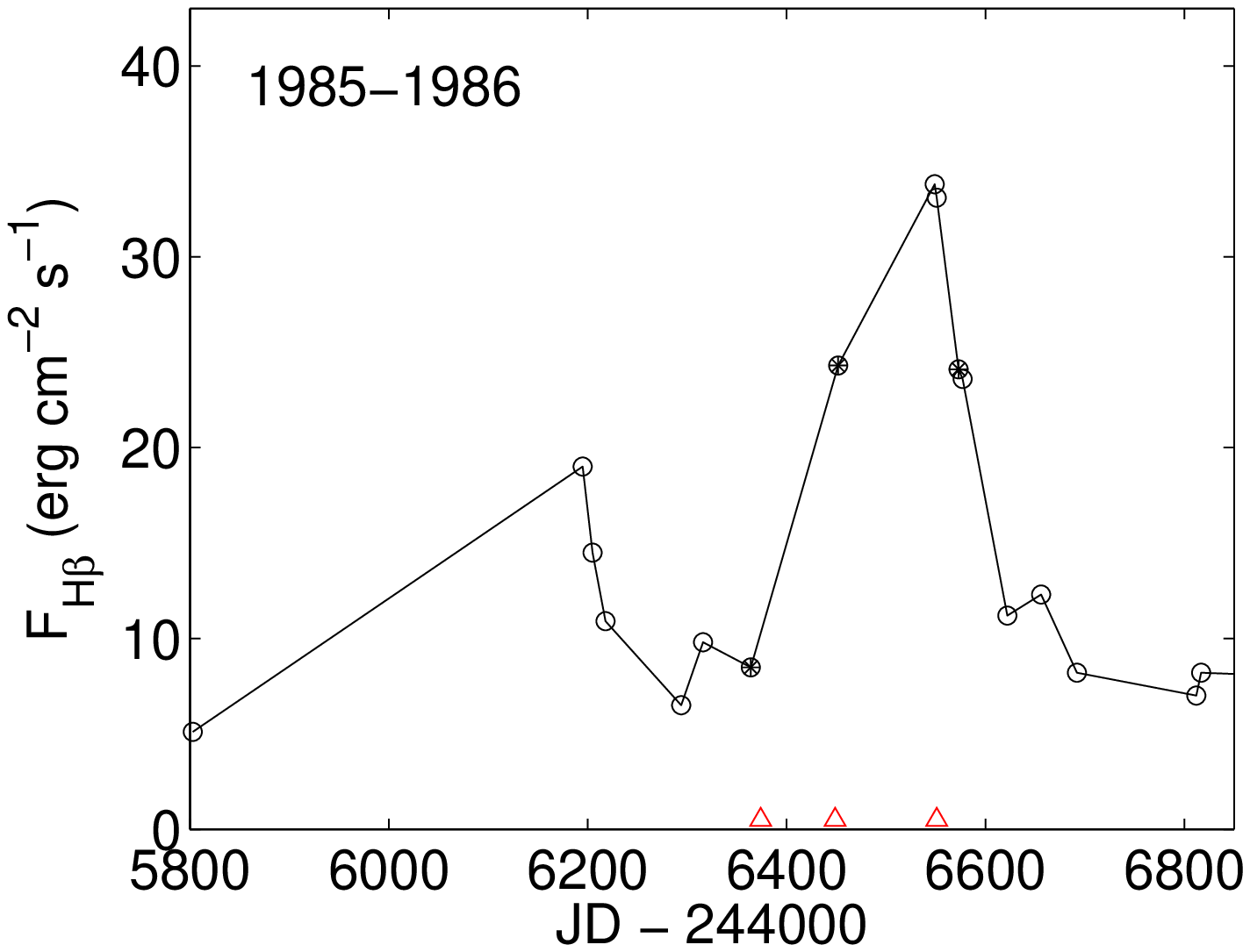}
\includegraphics[width=0.35\textwidth]{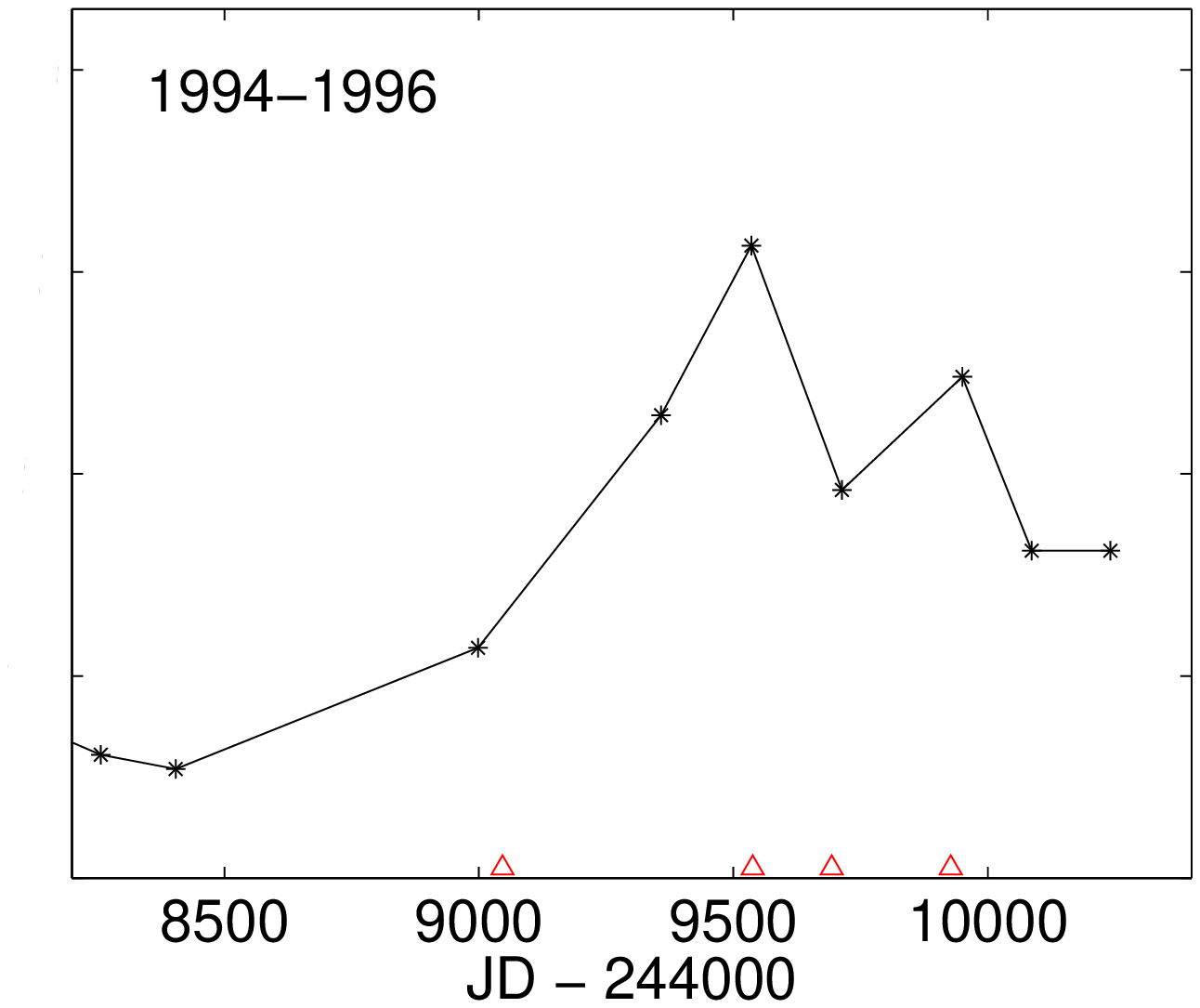}
\caption{Top diagram: the U and V  light curves of AG Dra from 1977  
adapted from Skopal et al. (2009, fig.1, left diagram). Bottom  diagram: variation of 
the strongest optical lines adapted from GVIG99, fig. 4. 
The red  triangles indicate the JD selected for modeling (see text).
\label{fig:lightc}}
\end{center}
 \end{figure*}

The SED  reveals the  relative importance of the different contributions to the
continuum (the  radiation from the stars, the bremsstrahlung from the nebulae, and the
reprocessed radiation from dust grains). Generally, in SSs,  the SED shows that  different nebulae
contribute to the spectra. Their  relative  contributions must be consistently accounted for
also in the calculations of the line ratios (Contini et al. 2009b) leading to
an iterative calculation process  which stops when  the calculated
spectra  and the data are well tuned.

The  data are  observed at Earth, while the calculations refer to the nebula.
By adjusting the calculated  to the observed absolute continuum
by  the factor $\eta$$\sim$ $r^2/d^2$,  which depends on the radius r of the
nebula and  on the distance d to Earth,
some information about the location of the nebulae is obtained.

\subsection{UV and optical lines}

We refer to the UV and optical spectra  observed by GVIG99  between June 1979 and February 1996. 
To   get a first  estimate of the conditions in the  nebulae  at different epochs,
we  combined the spectra in the UV and in the optical range  observed at adjacent
days. This  allows to  refer also the UV line ratios to \Hb which 
 is one of the most significant lines.  The analysis of the line ratios reveals
both the physical conditions and the relative abundances of the elements throughout the nebulae.
We  thus  selected the UV  spectra at 8 epochs: 4-11-1985 (JD 2446374), 18-1-1986 (JD ..6449), 30-4-1986 (JD ..6551),
27-2-1993 (JD ..9046), 4-7-1994 (JD ..9538), 6-12-1994 (JD ..9693), 28-7-1995 (JD ..9927), and 14-2-1996 (JD 245128)
which    are  randomly chosen phases.
The selected dates are  indicated upon the  diagram presented  in Fig. 1 (bottom), where we report the profile of the 
\Hb flux observed during the 1985-1986 and 1994-1996 active phases by GVIG99 (fig. 4, bottom diagram) and 
by Mikolajewska (1995). 

Optical line observations  are given by GVIG99 (table 2).
The dates  closest to the 8 epochs chosen    for the  UV  observations
are: 25-10-1985, 21-1-1986, 22-5-1986, 10-1-1993, 1-7-1994, 26-12-1994, 19-8-1995, and 3-1-1996, respectively.

The observations (GVIG99) show that  
the spectral region includes strong emission lines
of hydrogen,  helium, and weaker lines e.g. FeII and NIII. During the active phases
the strongest emission lines, HeII 4686, \Hb and H$\gamma$,  display extended wings
(FWZI  $\leq$ 3000 \kms) whose intensity is generally 10\% of the emission peak (GVIG99).
Broad wings were observed in the HeII 1640 line during the 1981 bright phase.

The optical spectra contain  the weak forbidden [OIII] 4363  line which has a relatively low
critical density for collisional deexcitation ($\sim$10$^5$ \cm3) for O$^{2+}$. 
Moreover, the [OIII] 4363 line  is observed
even  if [OIII] 5007, 4959, which   are generally the strongest lines,  are absent.
We obtain therefore  an   upper limit on the density.
 The observed [OIII] 4363 line, although very weak,
indicates  a stratification of temperatures and densities of the emitting gas, in agreement with
the profile of the physical conditions downstream of  a shock-front.

The HeII 4686/HeI 5876 line ratio  gives information about  the  photoionization flux and 
the recombination  conditions of the gas. 

In the UV, a strong NV 1240 depends on both the shock velocity and the intensity of the
primary flux, while MgII 2800, which is a relatively strong  low-level  line 
 depends indirectly on the density, which speeds up the cooling rate throughout the nebula. 
Actually, MgII can be strong if the emitting nebula is geometrically thick and radiation bound.
The binary separation constrains the geometrical thickness of the radiative nebula, yielding 
 a lower limit on the density downstream.
The HeII 1640/OIII]1660 line ratio plays a key role in the choice of the parameters,
 the density in particular. 

We have run a grid of models for each spectrum, first   selecting the physical parameters
which best reproduce  the line ratios, then  changing the relative abundances as a last refined improvement,
particularly for CIV and MgII which are the only representative lines of C and Mg, respectively.

We  reproduced the observed line ratios as  close as possible considering
observational errors  and  modeling approximation.
A fit within a factor of 2 is generally accepted.
However,  at some epochs, the optical lines, particularly [OIII] 4363 and HeI 4471, are underestimated
by  the  models which explain the line ratios in the UV. This indicates that the emission from
the expanding  nebula, which is characterized by relatively low densities, must be added. 
Besides  stronger  [OIII] and HeI lines, this nebula  emits  strong OI and MgII  because
it is  extended enough   to contain a region of  recombined gas.

After a first rough  modeling of the different spectra we have realized that the
temperature of the WD  is rather  stable at  $\sim$1.6$\times$ 10$^5$ K  on the days considered,
so we iterated the modeling process  
adopting for all the models this temperature  in agreement with 
 Skopal et al. (2009) and 
Greiner et al. (1997)  results by a fit to ROSAT quiescence data. Piro et al. (1985)  derived a bb 
temperature of 200000 K from June 1985 EXOSAT observations 
and the Zanstra temperatures derived from the HeII/F(1340 \AA) was 109000$\pm$5400 K in quiescence
(GVIG99).

In the phases  close to  outbursts, both the HeII 1640/\Hb and HeII 4686/\Hb line ratios  
are very high. However,  HeII 1640/\Hb = 14.9 cannot be reproduced by 
the  coefficients  (recombination, dielectric recombination, etc.) of He and H  
adopted  in the calculations (Osterbrock 1988 and references therein).
Notice that these lines  are saturated during the outbursts so  it is difficult 
to determine the broad and  narrow line relative intensity.
The best fit to the line spectra is obtained by  adopting a relatively high shock velocity which
leads to extended regions of  gas  ionized to  high levels.  A high ionization
parameter  produces higher  ratios to \Hb of intermediate ionization level lines
(e.g. [OIII]/\Hb), while a 
 high pre-shock density  increases the   cooling rate downstream changing  the stratification 
of the ions  and enhancing  the low ionization level lines.

We present in Table 1 the results  of modeling and we compare them with the observational data.
The  dates  corresponding to the UV spectra are shown in column 1, 
followed by the observed  UV
 and  optical line ratios to \Hb. 
The observed line and continuum fluxes  were reported  and corrected
adopting E$_{B-V}$=0.05, by GVIG99.
In Table 1, next to each   row  containing the  observed line ratios  the calculated  ones are shown.
Models m1-m8 which  provide the best fit to the data   appear in the first column of Table 1.
 Model m$_{exp}$ corresponds to the expanding nebula.

Models m1-m8 and m$_{exp}$ are described in Table 2, in terms of the physical conditions
in the emitting nebulae, selected  as input parameters  in  the SUMA calculation process. 
They lead to the best approximation of calculated to observed line ratios.
The first column of Table 2   shows   the  parameters  which  vary from
epoch to epoch, while those which were  found  roughly constant are  described in the
bottom of the table.
The shock velocity \Vs, the pre-shock density \n0, and the pre-shock magnetic field \B0
refer to the shock, while the color temperature of the hot star  \Ts and the ionization
parameter $U$ refer to the photoionizing flux from the WD. The geometrical thickness of the
emitting nebula $D$  determines whether  the model is radiation  or matter bound.
The dust-to-gas ratio ($d/g$) is crucial to explain the SED in the IR. The
relative abundances to H of the elements appear in rows 6-9 of Table 2.
The absolute \Hb flux calculated at the nebula is also given in Table 2, because it allows
to  calculate the absolute fluxes of all the lines.
A magnetic field \B0 = 10$^{-3}$ gauss is adopted for all models.

The calculated line ratios presented in Table 1 refer to the reverse shock between the stars (Sect. 2).
Therefore, the shock-front edge of the emitting nebula downstream
corresponds to the edge of the nebula reached by the photoionization flux from the WD. 
The velocity of the reverse shock  is generally considered approximately  stable.
A previous analysis (e.g. Contini \& Formiggini 2003) concludes that the reverse shock
between the stars is  actually a standing shock, although disrupted by  R-T  and K-H instabilities.

The values of the input parameters  which appear in columns 2-12  were selected because they 
explain the observational data. They will be considered as   $results$ of our  modeling.

In order to understand the trend of the line ratios, we show
the  profile of the electron temperature T$_e$ and electron density N$_e$
downstream, as well as that of the most significant ions, as a function of the distance from the
shock-front in Fig. 2 for  the nebulae relative to models m1, m5, and m5$_b$ (see Sect. 2.3).
Recall that the density downstream is higher than the pre-shock density by a factor
which depends on \Vs (besides the factor of 4 determined by the adiabatic jump
at the shock-front) and \B0. The physical conditions downstream follow the
recombination and cooling rates,  so  lines from different ionization levels
are emitted from the gas at different  distances from the shock-front.

The temperature of the gas is maximum in the immediate post-shock region 
T$\sim$1.5 10$^5$ ((V$_s$/(100 \kms))$^2$, then it decreases following the cooling rate  determined in each slab by
free-free, free-bound and line emission rates. At a temperature  $\leq$  10$^5$ K
recombination  produces  strong line emission. However, the temperature is maintained
at $\leq$ 10$^4$ K by the primary radiation from the WD and  the secondary radiation from hot gas slabs,
collisionally heated by the shock. Low ionization and neutral lines   emitted from
this region  weaken in geometrically thin clouds.

\begin{table*}
\caption{Comparison of calculated  with  observed  line ratios$^1$ to \Hb =1}
\flushleft
%\tiny 
\begin{tabular}{clllllllllclllll}\\  \hline  \hline
\        & NV  &OI  &OIV]+ &NIV]&  CIV  & HeII & OIII] & MgII & HeII&  H$\gamma$&[OIII] & HeI & HeIIn \\ 
\        &1240 &1340&1400  &1486&  1550 & 1640 & 1663  &2800  &3200 & 4340   & 4363  & 4471& 4686\\   \hline
\ 4/11/85& 1.16&0.165 &1.06&0.34& 1.14  &3.49  & 0.34  &0.33  &0.34 &0.36&  0.001   & 0.025&0.74     \\
\   m1   & 0.9 &0.18  &0.8+&0.6 & 1.47  &4.7   & 0.4   &0.32  &0.28 &0.37& 0.0014   &0.026 &0.65  \\
\ 18/1/86& 2.078&0.   &0.88&0.37&2.95   &14.66 &0.26   &0.2   &0.65 &0.51&0.009     &0.049 &1.4    \\
\   m2   &2.56   & 0.  &1.1&0.23&4.3    &7.    &0.2    &0.07  &0.4  &0.45& 0.       &0.07  &0.85  \\
\   m2$_b$&0.4   & 0.  &0. &0.  &0.2    &8.9   & 0.    &0.    &0.53 &0.49& 0.       &0.    &0.1   \\
\  m$_{exp}$&0.002&0.15&0.02&0.02&0.064&1.88   &0.23   &2.44  &0.11 &0.4 &0.04      &0.11 &0.27   \\
\ 30/4/86&2.39& 0.2     &1.26&0.28&2.26&4.     &0.253   &0.0  &0.33&0.44&0.006     &0.045 &0.64  \\
\   m3   &2.2   &0.1  & 1.8 &0.12&1.9  & 5.8   &0.26   &0.24  &0.35 &0.43&1e-4      &0.01  &0.78  \\
\ 27/2/93&0.78  &0.001& 0.35& -   &0.58& 2.97  &0.175  &0.184 &0.44 &0.35&0.001     &0.036 &0.526  \\
\   m4   &0.6   &0.001& 0.32&0.035&0.69&3.     &0.18   &0.17  &0.20 &0.46&0.        &0.03  &0.42  \\
\ 4/7/94 &3.9   &0.0  &3.35 &0.72 &2. & 10.  &0.63   &0.112 &0.473&0.38&0.        &0.093 &0.86   \\
\   m5   &3.63  &0.0  &3.7  &0.5  &3.  & 6.6   &0.5    &0.12  &0.40 &0.43&4e-4      &0.004 &0.86  \\
\ m5$_b$ &0.2   &0.0   &0.17&0.04 &16.7& 5.5   &0.2    &2.5   &0.33 &0.45&1.e-4     &0.015 &0.75  \\   
\ 6/12/94&1.24  &0.125&1.474&0.161&1.55&6.95   &0.4    &0.13  &0.35 &0.39&0.011     &0.049 &0.64   \\
\   m6   &1.1   &0.01 &1.6  &0.35 &1.53&6.     &0.7    &0.13  &0.3  &0.44&0.001      &0.01&0.78   \\
\ m6$_b$ &0.06  &0.005&0.05 &0.014& 6. &3.7    &0.11   &4.6   &0.22 &0.46&0.         &0.026&0.53  \\
\ 28/7/95& 1.5  &0.42 &1.61 &0.58 &2.73&8.85   &0.5    &0.185 &0.274&0.42&0.017     &0.06 &0.63    \\
\   m7  & 1.88   &0.   &2.   &1.1  &2.1 &7.     &0.6    &0.19  &0.42&0.44&0.004     &0.004&0.89    \\
\ 14/2/96&0.52  &0.1  &0.66 &0.48 &1.14&3.57   &0.376   &0.086&0.12 &0.34&0.015     &0.049&0.53    \\
\  m8   &0.5   &0.1 &0.7  &0.46 &1.58&4.6    &0.46   &0.054 &0.27 &0.39&0.02      &0.03 &0.64   \\     
%\  m$_{exp}$&0.002&0.15&0.02&0.02&0.064&1.88   &0.23   &2.44  &0.11 &         &0.4 &0.04      &0.11 &0.27    \\ 
\hline

\end{tabular}

$^1$ Gonz\'{a}lez-Riestra et al. (1999, Tables 1 and 2)

\end{table*}

\begin{table*}
\caption{Description of the models}
\flushleft
%\tiny 
\begin{tabular}{llllllllllllll}\\ \hline  \hline
\ model  &    m1  &    m2  &m2$_{b}$&  m3   &    m4  &   m5 & m5$_{b}$ & m6   & m6$_{b}$   & m7    &m7$_{b}$ &   m8& m$_{exp}$\\ \hline     
\ \Vs & 130 & 100    &1000  &  150    &   100  &  140  &700      & 120  & 700        &180    &700      & 120 &100\\
\ \n0 & 5.e8&2.4e10  &3.e7  &8.e9    &    3.e9 &  2.e9 &5.e8     & 1.e9 & 5.e8       &5.e7   & 5.e8    &3.e7 &3.e6 \\
\  $U$      & 2.5  & 1     &100   &0.1     &    0.5  &  0.2   & 5       &0.06  &  5         & 0.4   & 5       &2    &0.1 \\
\ $D$  & 2.6e13&7.8e10 &4.2e13&5.5e7   & 6.e11   & 8.8e8 &4.5e11   &6.e8  &4.6e11      &3.7e10  &4.6e11  &4.1e14&1.e14 \\
\ \Hb & 2.2e7&2.8e8  &2.2e4 &1.5e5   &  9.2e7 &2.8e4   & 9.4e6   &2.5e4 & 3.0e7      &2.6e3   &3.0e7   &3.8e6 &5.9e3 \\ 
\ C/H     & 1.3e-5 &8.0e-6 &3.3e-4&1.3e-5  &  1.2e-5 &1.3e-5&3.3e-4    &1.7e-5&3.3e-4       &1.e-5    &3.3e-4 &1.3e-5&1.3e-5 \\
\ N/H     & 4.e-5 &2.1e-5  & 9.1e-5&2.1e-5 &    8.1e-5& 2.1e-5&9.1e-5  &2.1e-5 &9.1e-5      &2.1e-5   &9.1e-5 &2.1e-5&2.1e-5\\
\ O/H     & 2.e-4 &1.6e-4  &6.6e-4 &1.6e-4 &   1.6e-4&1.3e-4&6.6e-4   &1.6e-4 &6.6e-4      & 1.e-4   &6.6e-4  &1.6e-4&1.6e-4\\
\ Mg/H    & 2.e-5  &2.e-5 & 3.e-5  &3.e-6   & 2.0e-6 & 1.5e-5&3.e-5   &3.e-6  &3.e-5       & 4.e-5   &3.e-5  & 3.e-6&3.e-6\\
\ r       & 8.4e13 & 2.4e13 &4.2e12&7.5e14&2.6e13  &  9.4e14& 1.8e11 &1.7e15 &  1.88e11   & 1.9e15  &1.3e11 &1.7e14&2.4e14\\ 
\ R       & 5.8e12 & 1.6e12 &6.7e12 &3.4e12& 4.8e12  & 5.6e12 &4.e11   &1.6e13 &2.0e11      & 2.e13   &5.3e11  &2.8e13&2.e14 \\
\ $\eta$  & -15.8  & -17.   & -18.5 & -14. & -16.9  & -13.5  & -21.3  & -13.  & -21.2      & -12.9   & -21.5  & -15.3& -15. \\
\hline
\end{tabular}
\flushleft
\Vs is in \kms, \n0 in \cm3, $D$ in cm, and \Hb is in \erg. 

For all the models \B0=10$^{-3}$ gauss, \Ts=160000 K, He/H = 0.1, Ne/H= 8.3 10$^{-5}$, Si/H=3.3 10$^{-6}$,
S/H= 1.6 10$^{-6}$, A/H = 6.3 10$^{-6}$, Fe/H = 6.2 10$^{-6}$, and $d/g$=10$^{-14}$ by number.
\end{table*}

\begin{figure*}
\begin{center}
\includegraphics[width=0.32\textwidth]{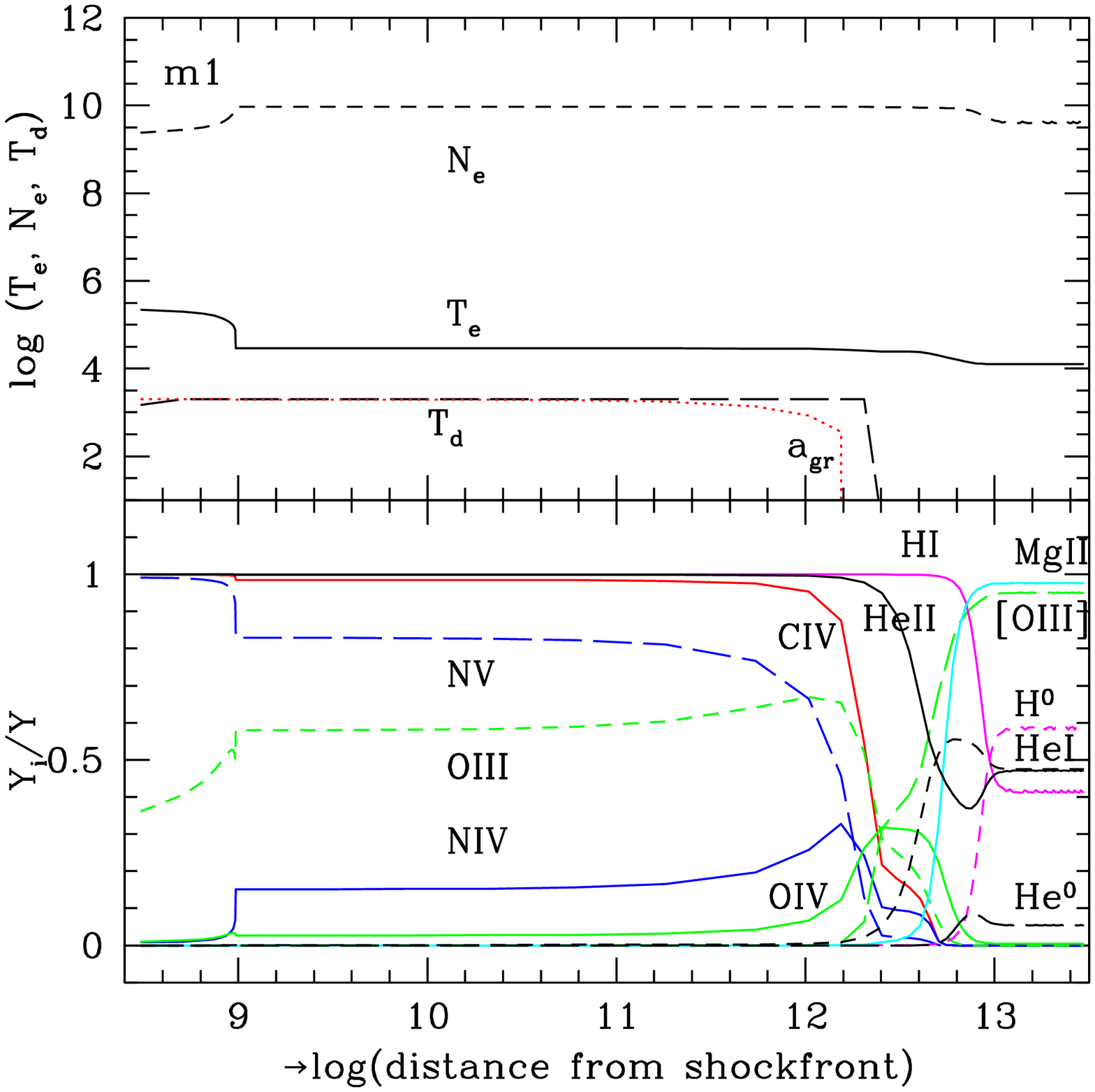}
\includegraphics[width=0.32\textwidth]{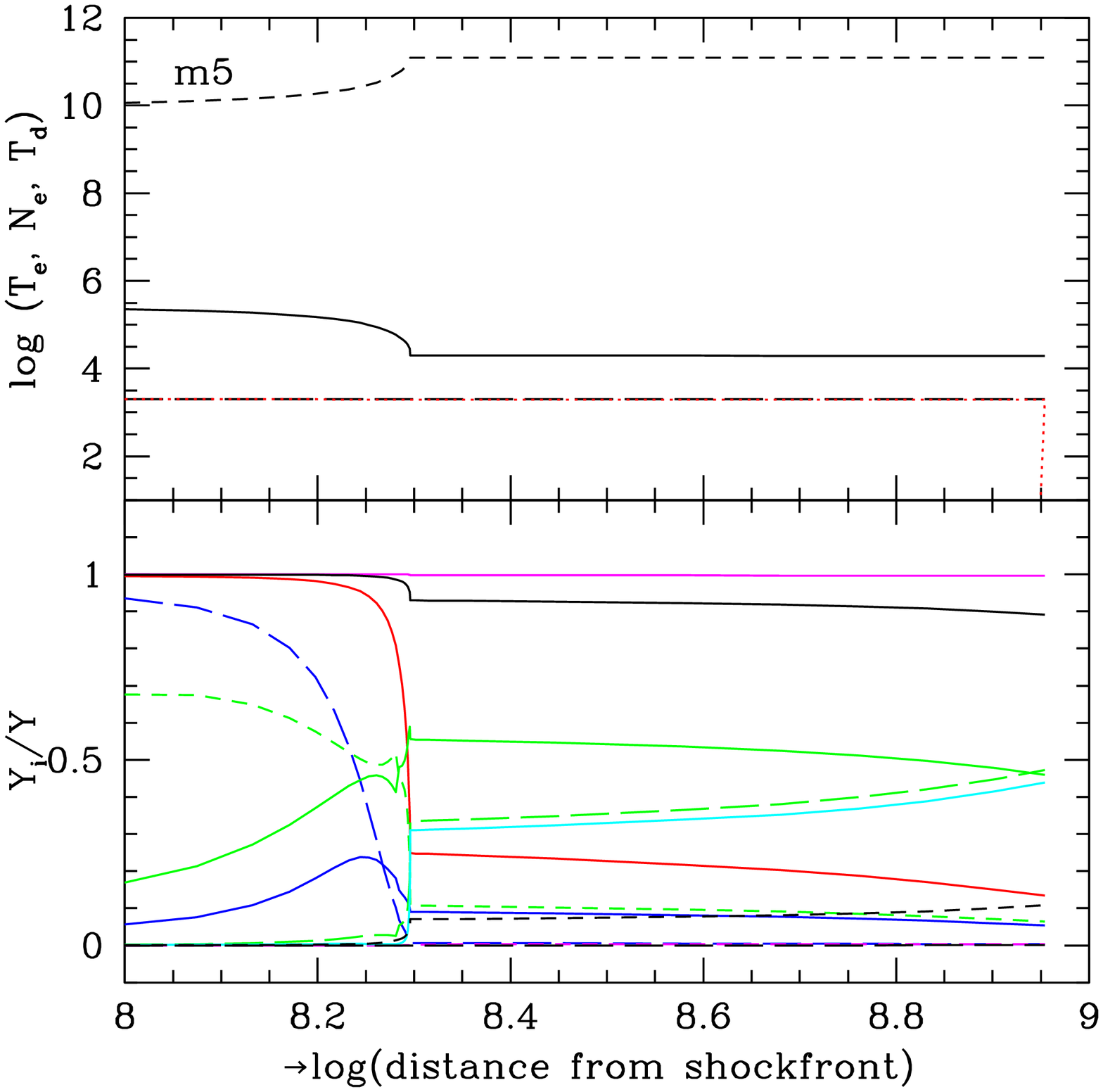}
\includegraphics[width=0.32\textwidth]{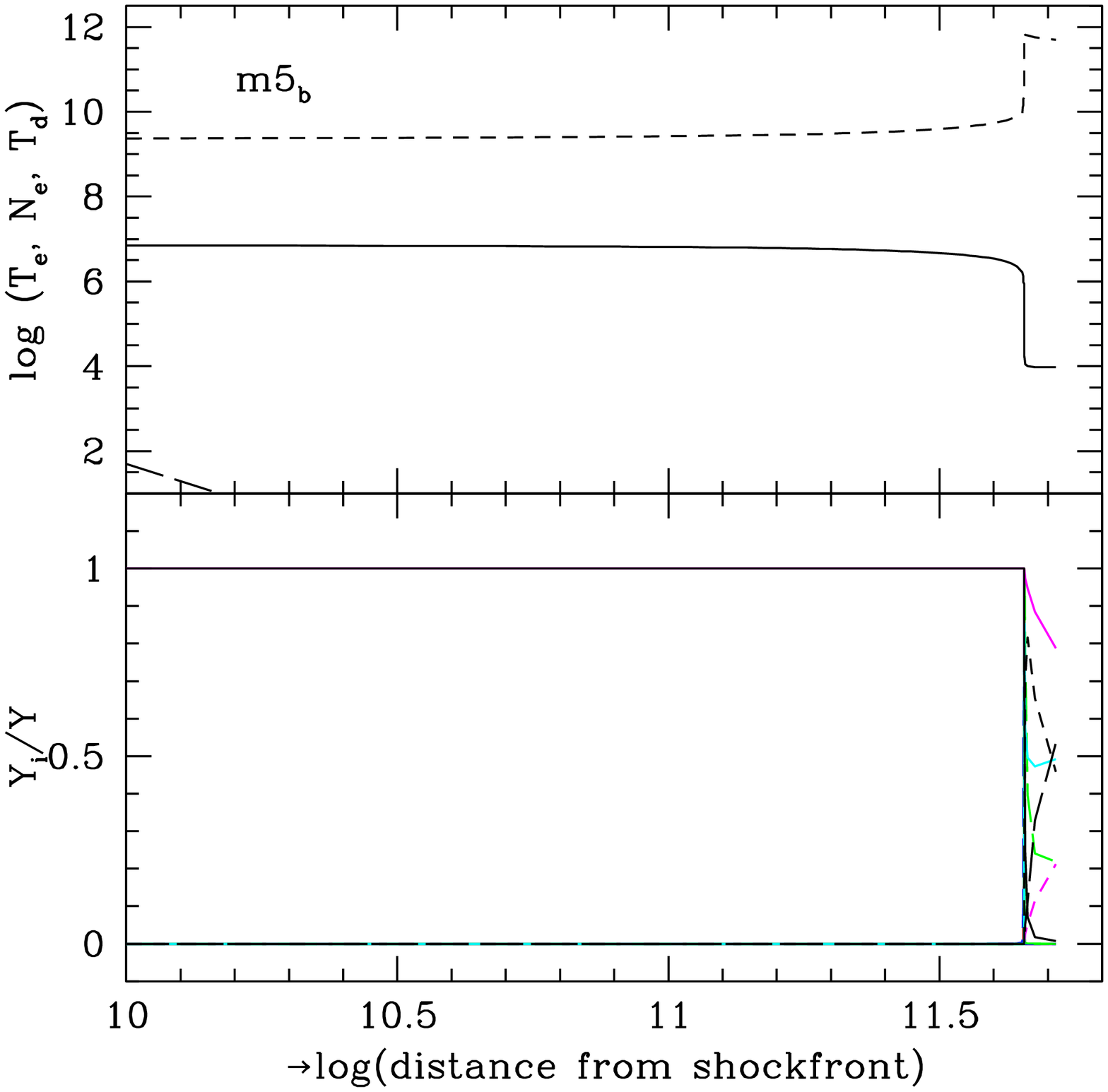}
\caption{Top panels: the profiles of the electron temperature, electron density and the dust temperature in the 
regions downstream emitting the different
lines for models m1 (left diagram), m5 (middle) and the model calculated with \Vs=700 \kms  (right diagram). 
The dotted red line refers to the grain radius a$_{gr}$
in \AA.
Bottom panels: the profiles of  the most significant ion relative fractions.
\label{fig:ions}}
\end{center}
\end{figure*}

\subsection{Broad lines}

GVIG99  present the broad component flux of HeII 4686 and \Hb  corrected for extinction in their table 2.
We have constrained the model  assuming \Vs  between 700 and 3000 \kms,
\n0 $>$ 10$^7$ \cm3, $U$ is derived from the HeII$^b$/H${\beta}^b$ ratio, and  the radius of the nebula  within the 
AG Dra system (Sect. 4.2).
We have found that \Vs=1000 \kms, \n0=3 10$^7$ \cm3, and $U$=100  gives HeII$^b$/H$_{\beta}^b$ =1.
The observed ratio is 1.6.
The  H${\beta}^n$/H${\beta}^b$ line ratio, calculated by this model, results in 15.2 at Earth  in agreement 
with GVIG99 observations. 
Models m2$_b$,  m5$_b$, etc.,  are calculated adopting solar abundances because sputtering destroys the grain at 
such high \Vs (Fig. 2).

In the following active phase, starting at day 4/7/1994, we found that a model with \Vs=700 \kms,
\n0=5.10$^8$ \cm3 and $U$=5 reproduces the observed  HeII$^b$/H${\beta}^b$ $\sim$0.8 for  $D$= 4.55 10$^{11}$ cm
and  $\sim$0.5 for $D$= 4.6 10$^{11}$ cm, as observed   on 6/12/1994, and 28/7/95, respectively. 
The HeII/\Hb  ratios in fact decrease  with  temperature
 at  larger geometrical thickness.
We do not have  enough data to further constrain the input parameters of models  m6$_b$ and m7$_b$  and they are therefore 
 identical.
These models contribute  a  weak broad component  to  the CIV line profile, which is not
confirmed by the observations (GVIG99).  
Carbon could be included in  small graphite grains which are, however, 
completely sputtered at such relatively high shock velocities. Therefore,
 depletion of carbon  could be due to the presence of  CO molecules, the second most abundant species
in the ISM after H$_2$.

\subsection{The continuum SED}

 In Fig. 3 we  compare the observed 
continuum SED   with model calculations  at  each of the  selected epochs. 

The  analysis of the line spectra at each epoch 
 determines the bb temperature of the WD and the  characteristics of the  nebulae  downstream of the
network shock-fronts. 
The  analysis of the continuum SED reveals 
the  presence of dust   shells. The temperature of the RG  is approximated by a bb flux.
We use an effective temperature of the  cool star 
T$_{eff}$=4300 K consistent with AG Dra's classification as an early K giant (Lutz et al. 1987)

To constrain the continuum
we have used the  two  data observed at 1340 and 2880 \AA~
by GVIG99. Indeed two points are  too few to give a somehow complete 
information, however, the two wavelengths reported are  
 strategic  enough  because one (2880 \AA) belongs
to the   bremsstrahlung  from the nebulae and the other (1340 \AA) often
to the  bb flux from the WD.

The nebulae are  located in the  circumstellar region
between the stars and/or in the circumbinary regions. The radius r of the nebulae
 can be obtained  by  comparing the calculated  with the observed absolute
fluxes. They are given in the bottom of Table 2 providing  a picture of the system  on a large scale
(Sect. 4.2).

In  Fig. 3   we     present the observations obtained in the X-ray and UV ranges 
by XMM and FUSE, respectively by Skopal et al. (2009, fig. 2, left panels).
The data in the radio come from Mikolajewska (2002). We have inserted them in the 
spectrum referring to 30/4/1986, even if the  epochs  were not exactly the same, to constrain
the bremsstrahlung in the radio range.
Actually,  Angeloni et al. (2010) claim
that radio variability also  throughout a relatively long period does not affect the main results of modeling.
The optical low-resolution spectrum comes from the Loiano observatory and the flux points are determined
by broad band optical UBVR$_C$L$_C$ and near-IR JHKLM photometry (Skopal et al. 2009).

\begin{figure*}
\begin{center}
\includegraphics[width=0.32\textwidth]{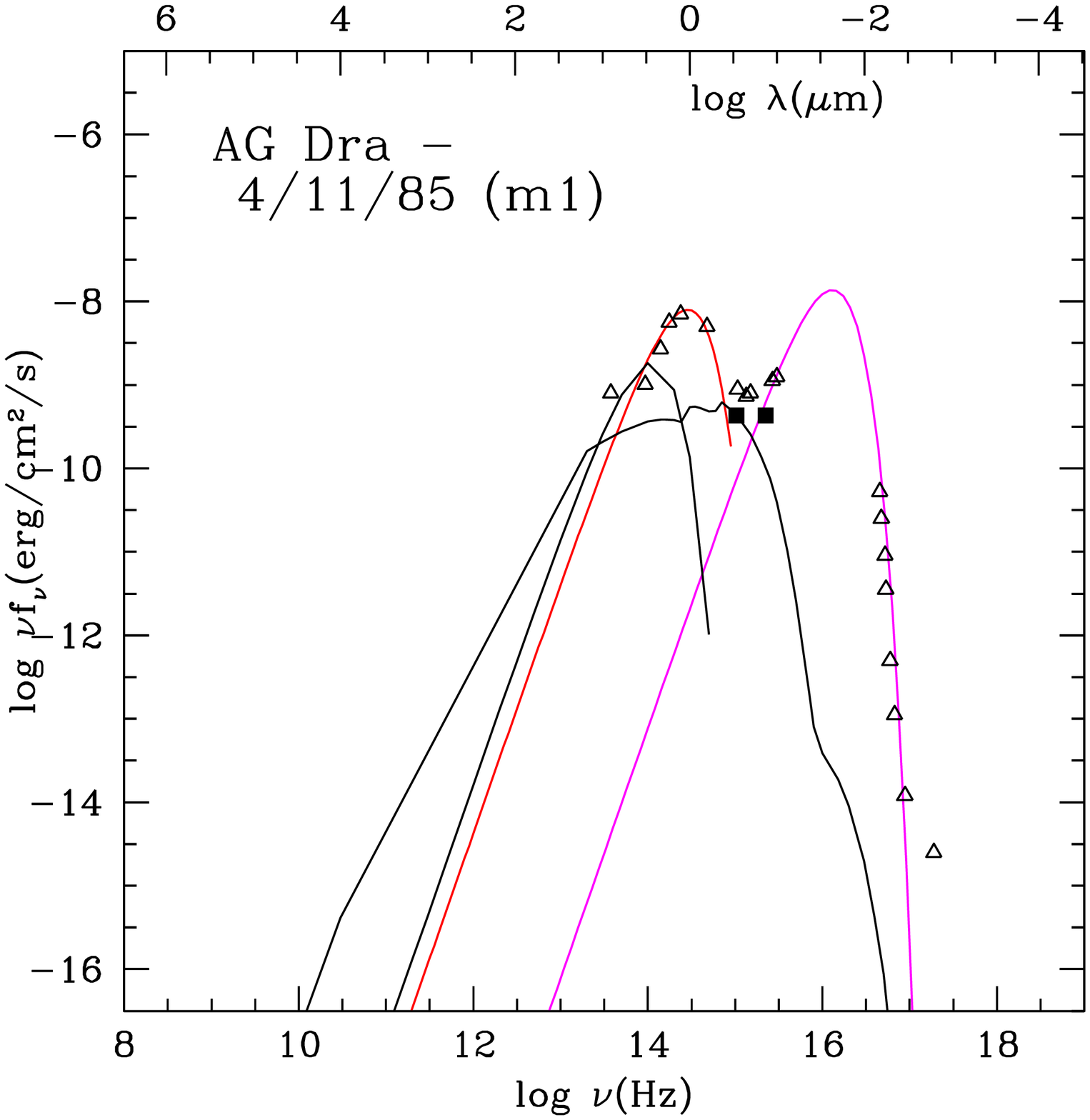}
\includegraphics[width=0.32\textwidth]{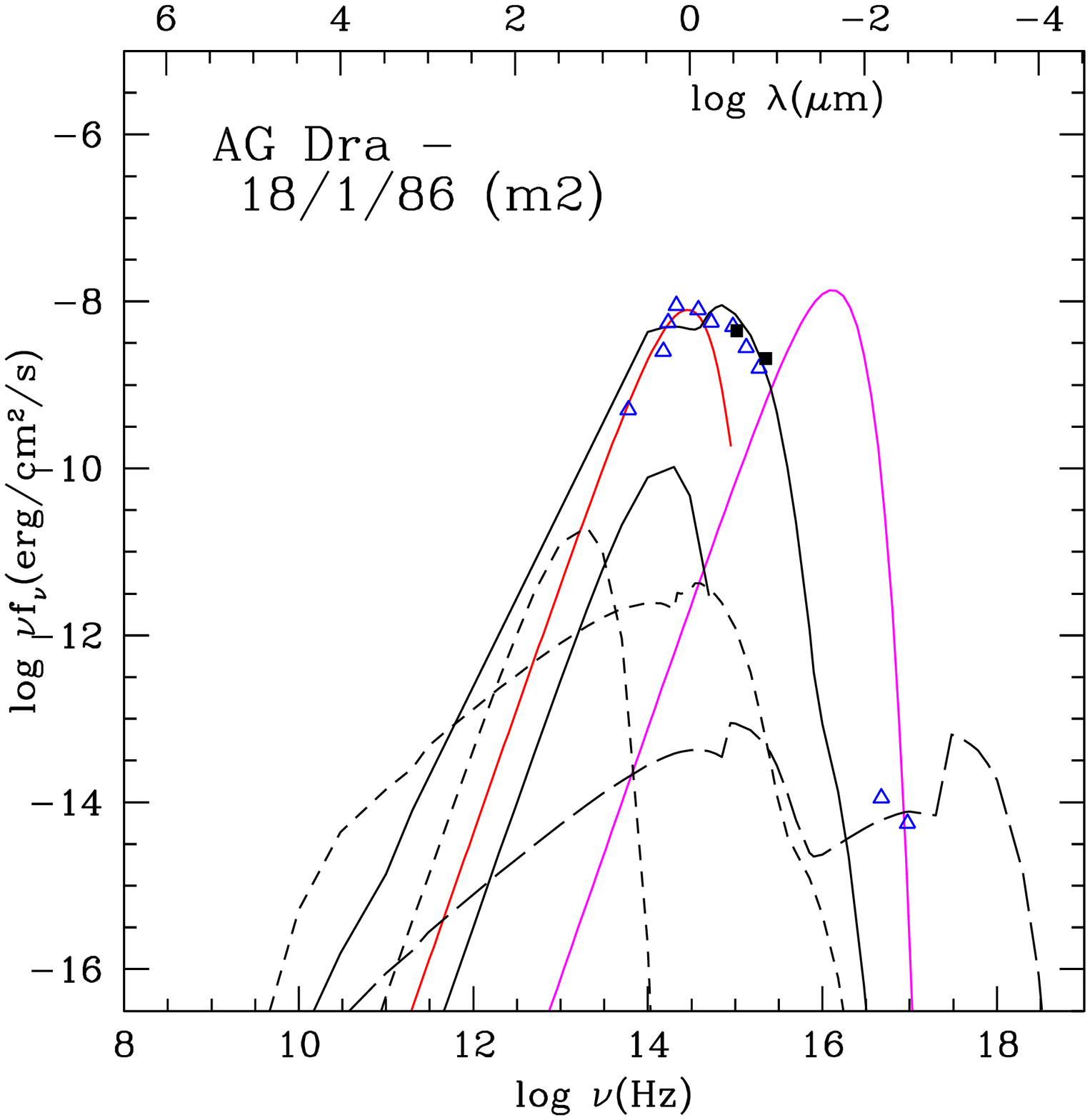}
\includegraphics[width=0.32\textwidth]{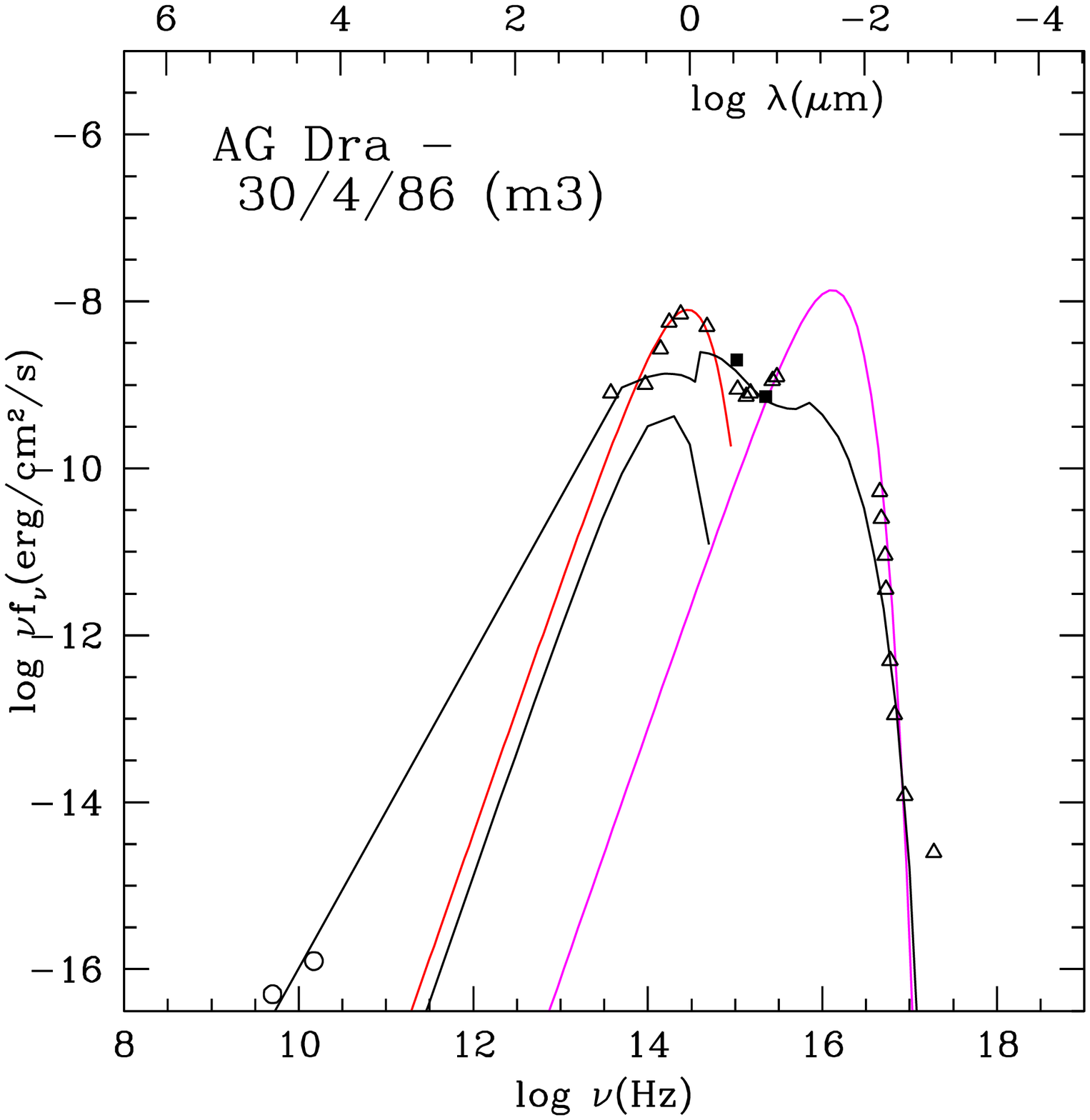}
\includegraphics[width=0.32\textwidth]{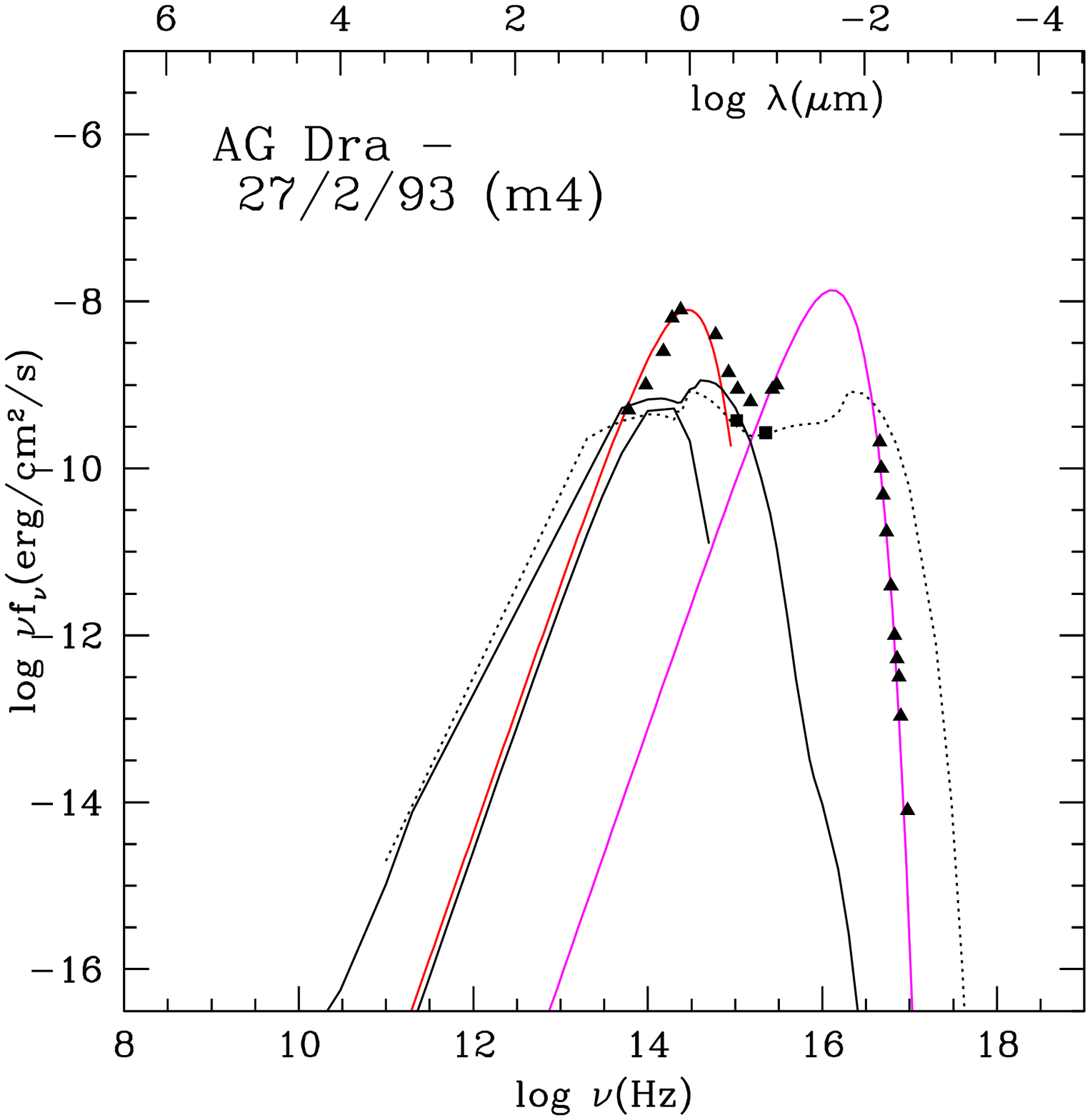}
\includegraphics[width=0.32\textwidth]{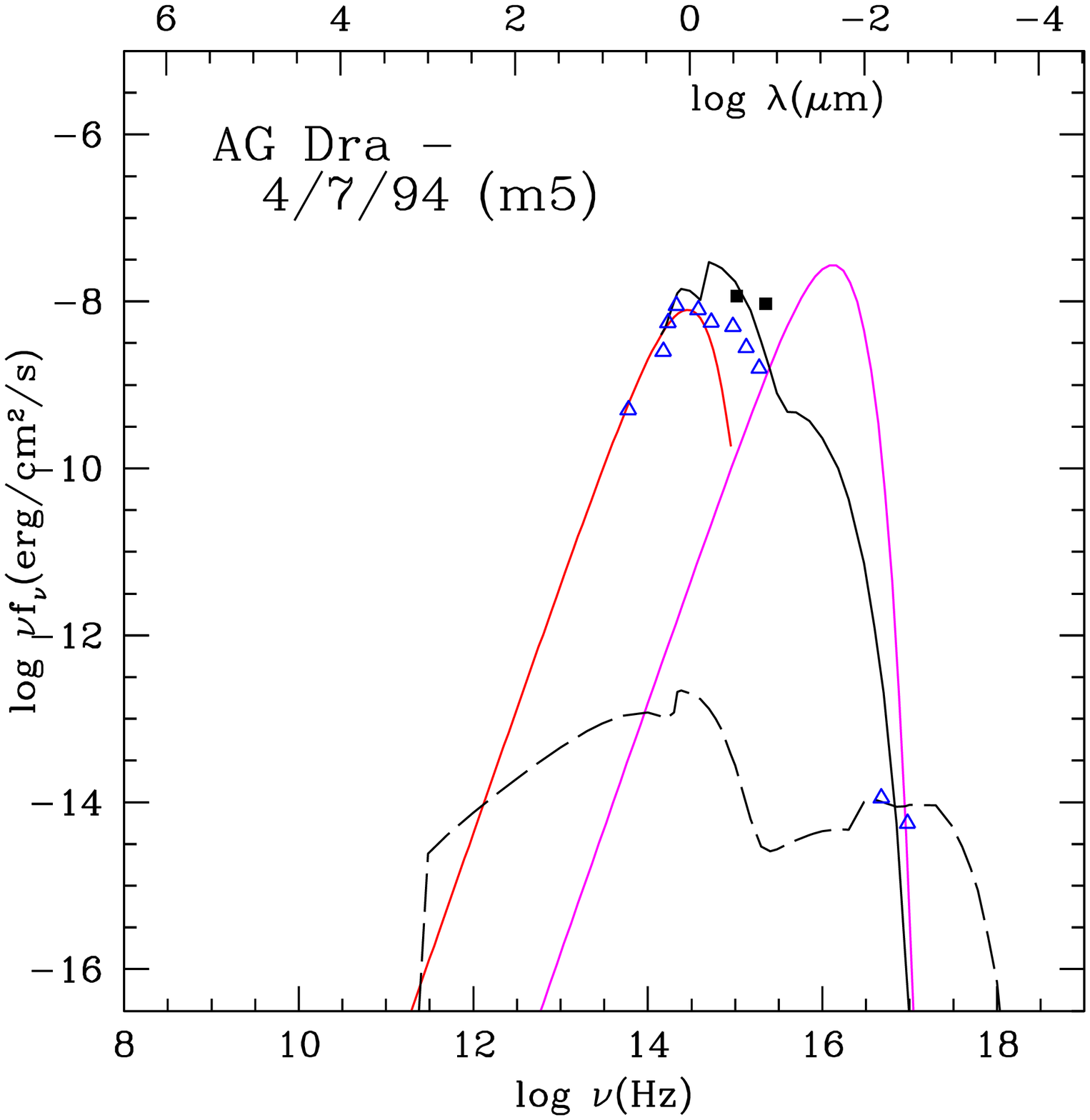}
\includegraphics[width=0.32\textwidth]{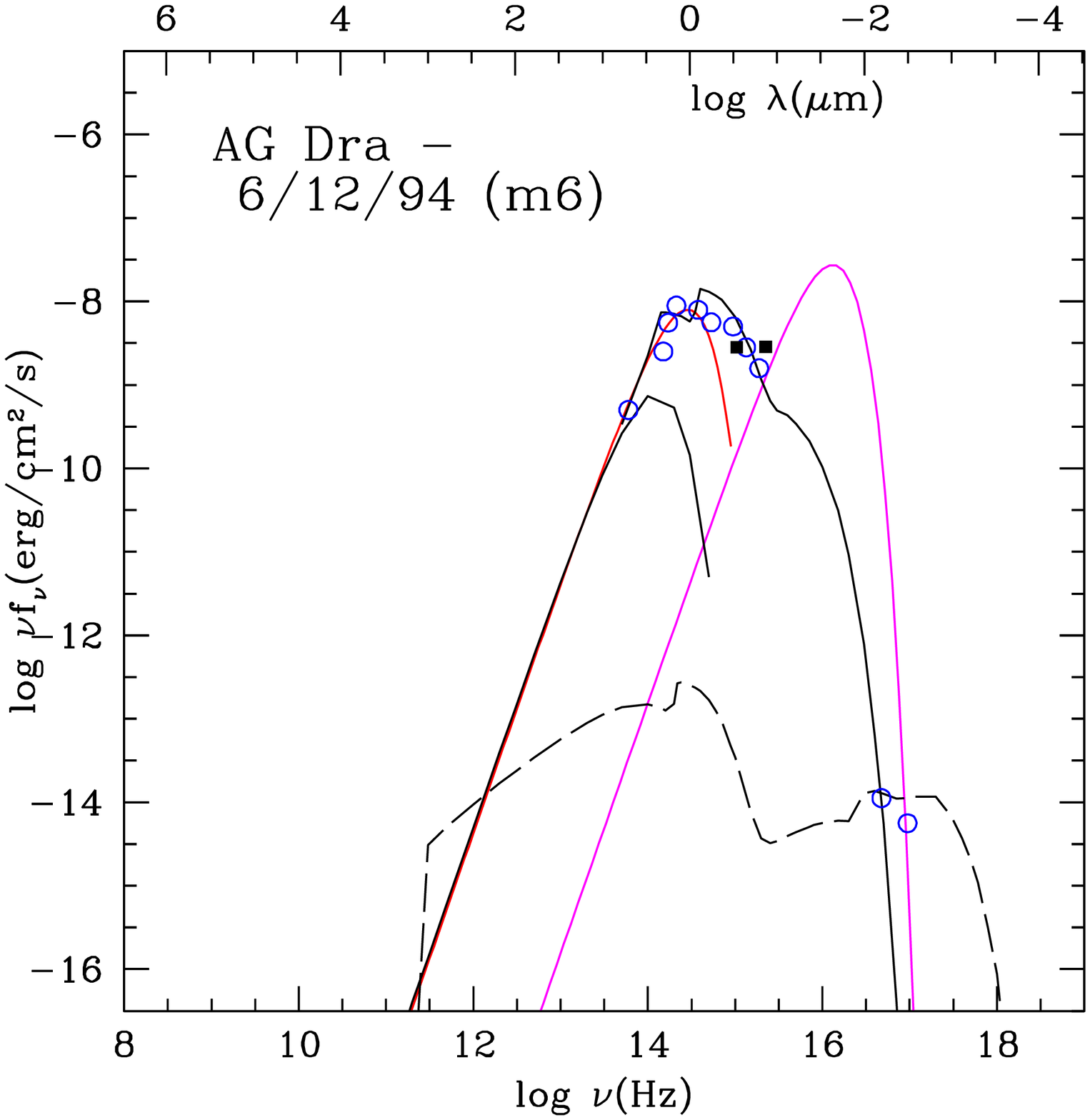}
\includegraphics[width=0.32\textwidth]{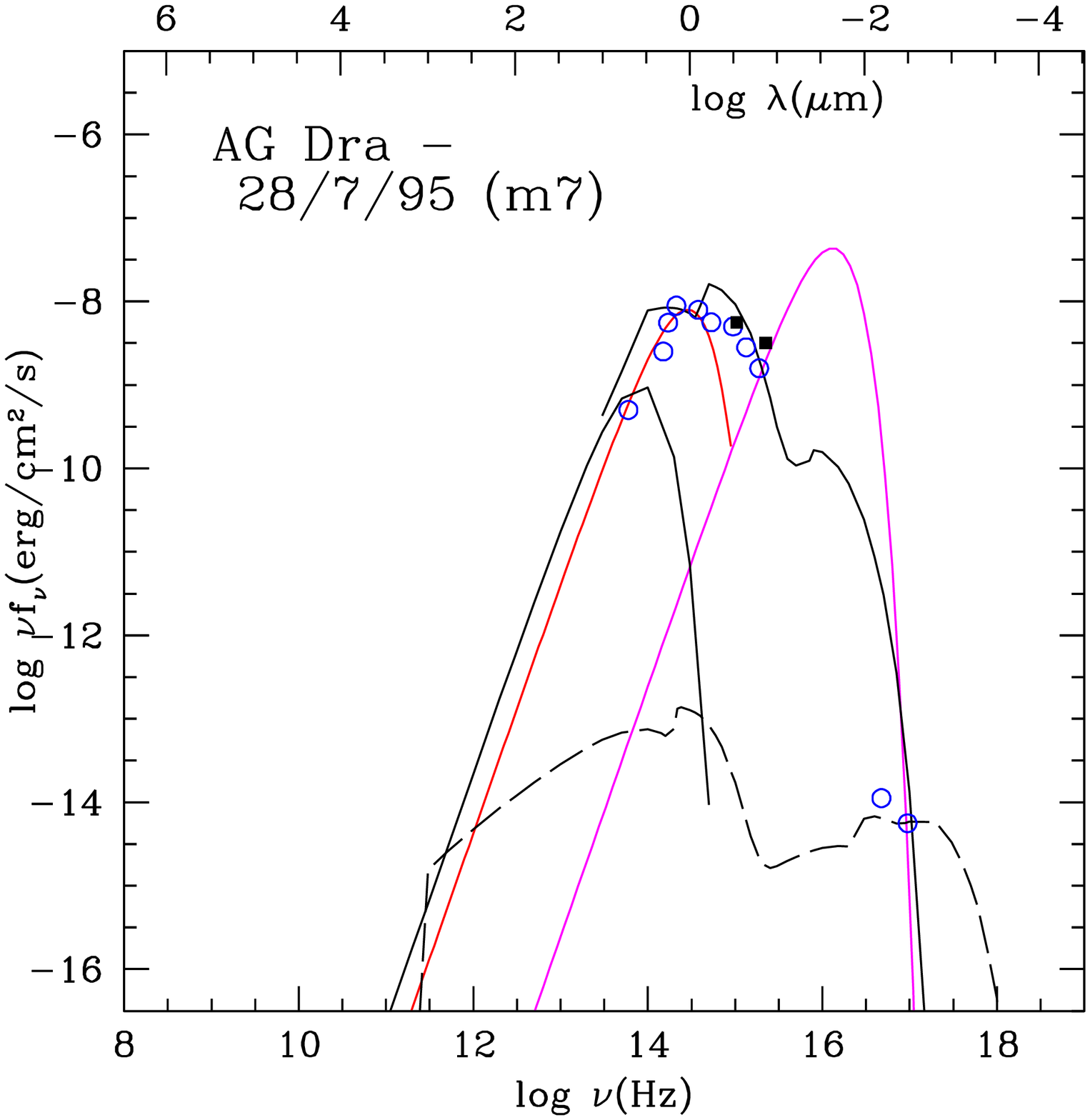}
\includegraphics[width=0.32\textwidth]{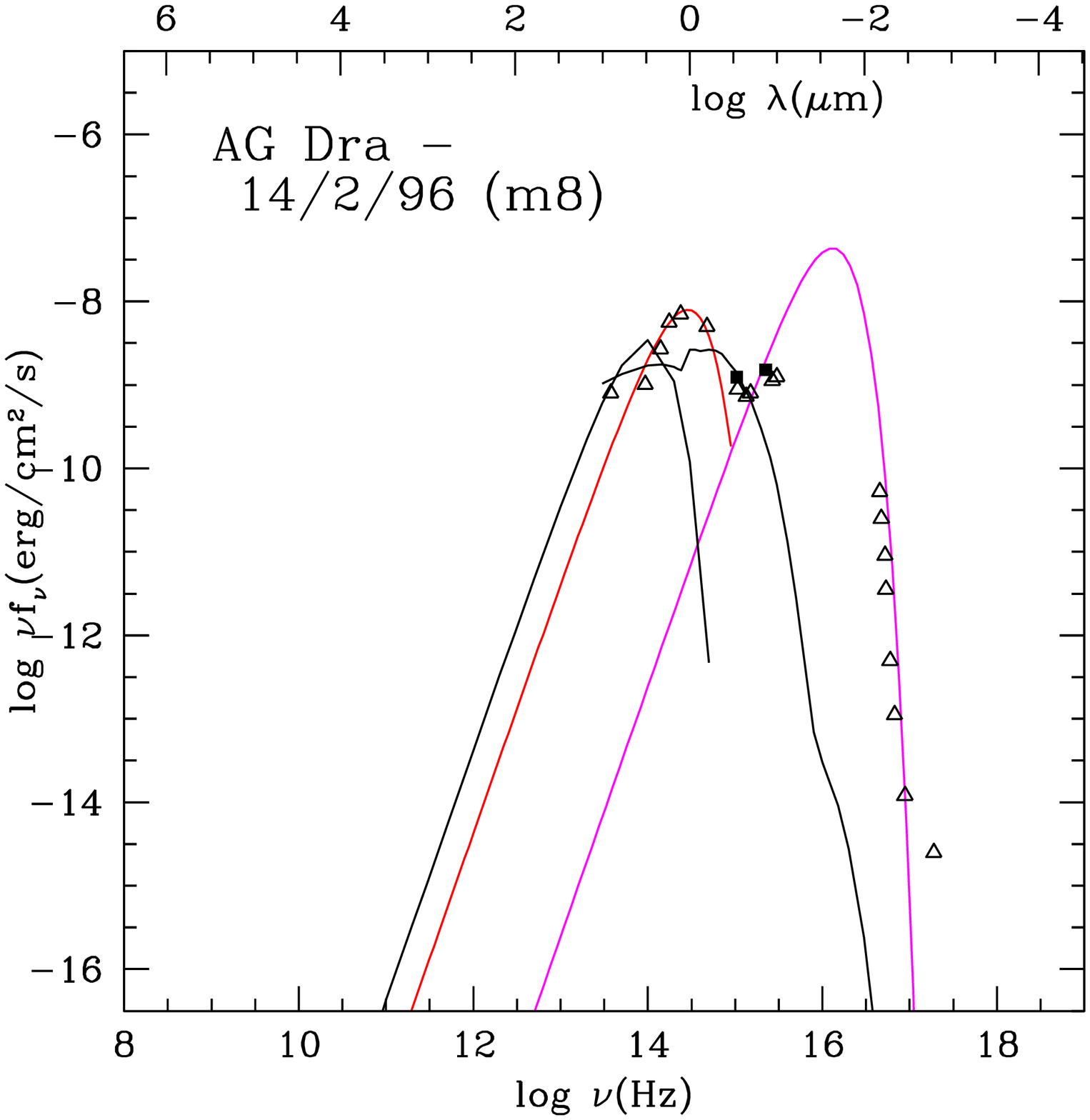}
\caption{The modeling at different phases. The observations by Skopal et al
(2009) are indicated by  triangles: open blue at burst, filled black  at quiescence,
open black at transition. Black filled squares are the continuum data from GVIG99.
Red solid line: the bb flux from the RG; solid magenta line: the bb flux from the WD;
solid black line:  the two curves represent the bremsstrahlung from the nebulae described in Table 2
and reprocessed radiation by dust; long-dashed
black line: the bremsstrahlung from the high velocity models; short-dashed black line:
the expanding nebula; dotted line: bremsstrahlung from a model calculated  by \Vs=280 \kms.
\label{fig:seds}}
\end{center}
\end{figure*}

\section{Results}

We have modeled the line and continuum spectra emitted from AG Dra at different phases
at  epochs corresponding to close UV and optical observation days. 
The results are presented in the following.

\subsection{Gas physical conditions}

The input parameters \Vs, \n0, and U which  provide the best fit
of the line ratios at different epochs are shown in   Fig. 4 (left diagram). 

 We  found low-velocity (lv) shocks with
 \Vs  between 100 and 180 \kms ,
in agreement with the FWHM of the UV high-level  line profiles (v=100-160 \kms), 
 e.g. MgVII 2510, 2629, SiVII 2147, etc., reported by
Young et al. (2006)  from  recent Hubble Space Telescope STIS observations.

The  high velocity  (hv) shocks  (\Vs = 700-1000 \kms)  
 explain  the broad wings in the Balmer  lines   observed close to the
outburst epochs. They are essential in modeling  the high He II 1640/\Hb~ line ratios
at the dates 18/1/1986 , 4/7/1994, and 28/7/1995.
The hv shocks accompany the blast wave from the explosion,  which  propagates radially,  except
towards the RG,   where  collision with the dense RG wind prevents  its expansion and  
creates  a complex network of shock-fronts.

The SEDs of the continuum emitted   from the nebulae downstream of the hv shock-fronts are
very  different from  those reported by Skopal et al. in quiescent and transition phases (Fig. 3).
In particular they  produce negligibly low UV-optical fluxes.
On the other hand the continuum SEDs  emitted from the  nebulae downstream of the lv  shock-fronts 
 between the stars can reproduce the trend of
the observations in the soft X ray range for  \Vs $\geq$ 150 \kms.
For \Vs  $>$ 180 \kms the X-rays can even overcome the bb flux from the WD, mimicking a higher WD temperature.
For example,  the SED calculated with \Vs=280 \kms in the transition phase
on 27/2/1993 is shown in Fig. 3.\\

The pre-shock densities are relatively high, leading to the low velocities accompanying  
the reverse shock between the stars in AG Dra (from conservation of mass at the shock-front:
n0v0=n1v1, where 0 refers to upstream and 1 to downstream conditions).
In  R Aqr the velocities of the reverse shock are   $\sim$100 \kms
and  the densities are $\sim$10$^5$ \cm3  (Contini \& Formiggini 2003).
The velocities of the reverse shock  inferred  from the IR line ratios
in D-type SS (Angeloni et al. 2007a) are $\sim$400-500 \kms. We wonder whether
dust and/or higher densities could be responsible for the deceleration of the reverse shock.\\

The input parameter $U$ varies between a maximum of  100 in the high velocity nebulae to a minimum
of 0.06  for model m6. A high $U$ indicates that the nebula is close to the hot source and/or
that there is no matter  obstructing the photoionizing flux  
(cfr. for Z And - Contini et al. 2010, in preparation).

\subsection{Radius of the nebulae}

The comparison  of the continuum fluxes calculated  at the nebula with those observed at Earth determines
the radius r of the nebula  by the factor $\eta$ (r$^2$= 10$^{-{\eta}}$ d$^2$ \ff, where \ff ~
is the filling factor, d the distance to Earth of AG Dra). Adopting d=2.5 Kpc  (GVIG99) and \ff=1 we obtain the 
radius upper limits 
which appear in row 10 of Table 2.
Considering  that the binary  separation is  2.3 10$^{13}$ cm (Leibowitz et al. 1985,
Tomov \& Tomova 2002)  \ff  is likely to be $\sim$0.1- 0.01.

We have  calculated the distance of the inner  edge of  the nebulae from the WD by
combining {\it F} (the number
of photons cm$^{-2}$ s$^{-1}$
reaching the nebula) with the ionization parameter $U$ calculated phenomenologically
by modeling the line spectra,

\noindent
 {\it F} (r$_{WD}$/R)$^2$ = $U$ n c, where r$_{WD}$ is the radius of the WD (0.08 R$_{\odot}$ GVIG99), 
 n the density and c the speed of light.
The resulting R are given in   row 11   of Table 2.

Model (m$_{exp}$) should  be added to the reverse shock  models  at the dates
30/4/1986, 4/7/1994/, and 28/7/1996,  increasing  the flux of the  low-ionization and neutral lines.
This model  represents the  nebula  downstream of the shock propagating out of the system (Contini et al. 2009c). 
The shock-front edge of the nebula
  is opposite  the edge reached by the flux from the WD.
  \n0 is lower and the nebula  is extended.
On the other hand, the nebulae downstream of the  hv shocks (700-1000 \kms) have a very small
radius because they represent matter ejected   at the outburst which had not yet  the time to
propagate  in the surroundings of the WD.

\begin{figure*}
\includegraphics[width=0.45\textwidth]{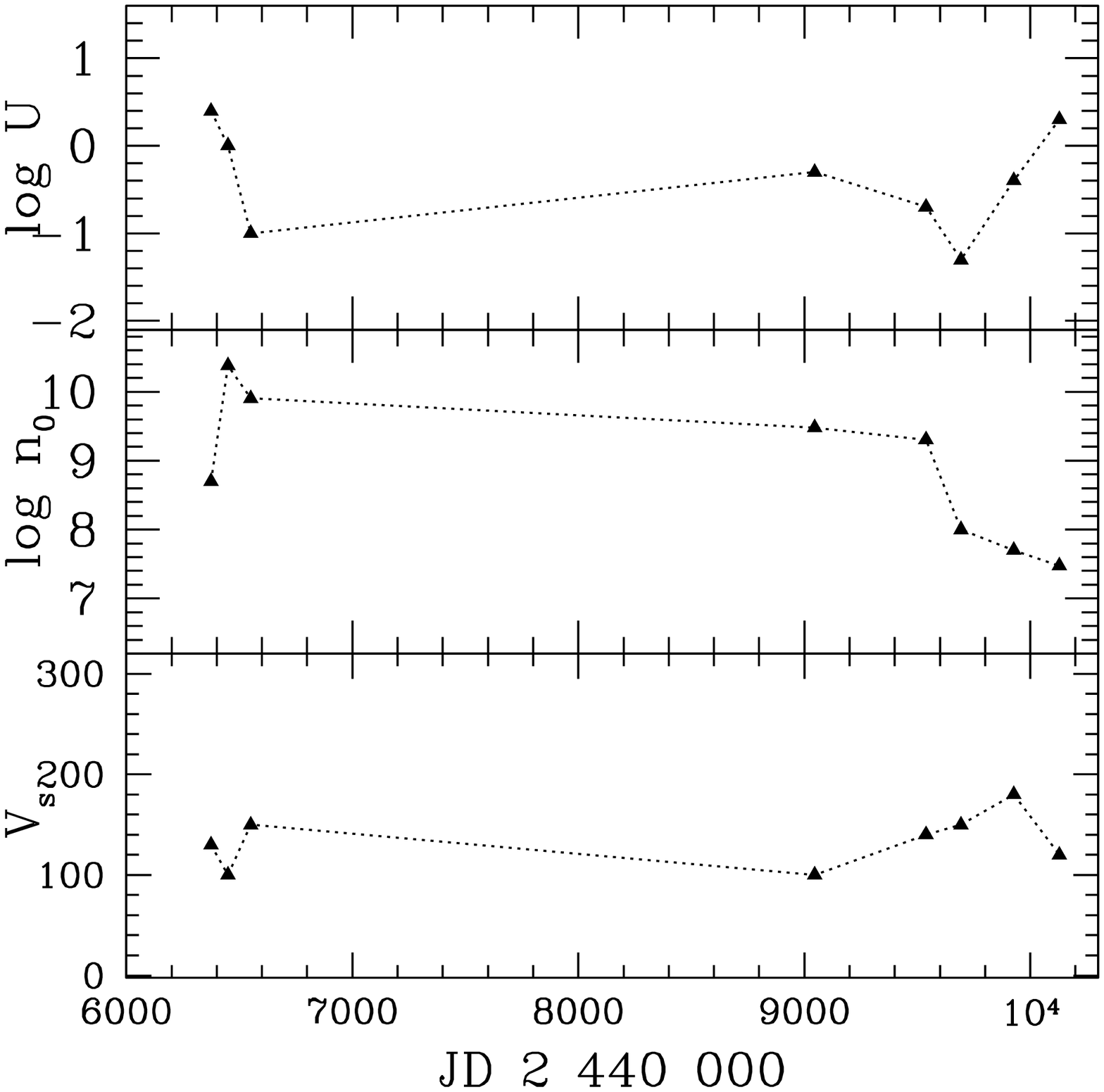}
\includegraphics[width=0.45\textwidth]{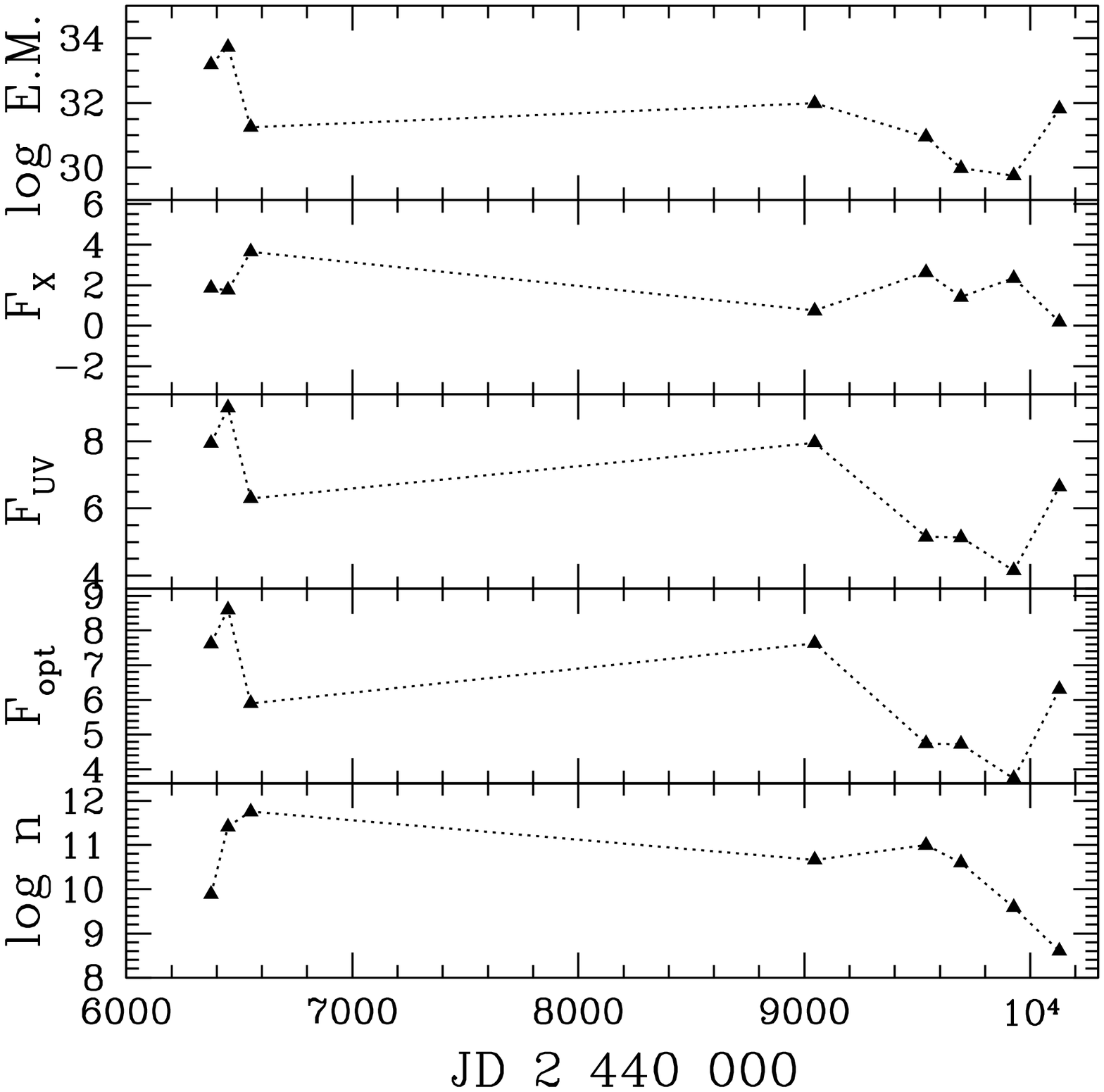}
\caption{The input parameters (left diagram), the post-shock density,  the fluxes
in the optical, UV, and X-ray domain, and E.M., calculated by the models (right diagram)}
%\label{fig:ions}}
\end{figure*}

\subsection{Integrated UV and soft X-ray fluxes  from the nebulae}

The input parameters  (Fig. 4 left) explain  the trend of the integrated fluxes (Fig. 4 right)
in the optical F$_{opt}$ (8267-3100 \AA),  UV F$_{UV}$ (3100-620 \AA), and in the soft X-ray range F$_{X}$
(0.1 -1.2 keV). 
 The maximum  gas densities downstream, i.e. the densities
which  are used to calculate the fluxes for each model are  shown in the right  diagram, as well as
the  E.M. (the emission measure $\Sigma$ $D$ n n$_e$). It can be noticed that
F$_{opt}$ and F$_{UV}$ follow the trend of $U$,
while  F$_X$ increases  with \Vs, because the maximum temperature
downstream is $\propto$ V$_s^2$.

A rough anti-correlation (Fig. 5) is  evident between  F$_X$ and F$_{UV}$  calculated
 at the nebula by  models m1 - m8, m2$_b$, and m5$_b$ .
The fluxes calculated in models m6$_b$ and m7$_b$  coincide with m5$_b$, therefore, they are
removed from Fig. 5 for the sake of clarity.
Recall that the higher \Vs the larger is the high temperature
zone within the nebula because recombination coefficients are low.
The cooling rate increases rapidly when the temperature of the nebula decreases below
10$^5$ K. At this  stage a strong compression followed by a rapid cooling rate ($\propto$ n$^2$)
 reduces the geometrical thickness of the UV emitting region (Fig. 2, diagram on the right)
and the UV flux is relatively low. So the SED of the continuum  from hv and lv shocks
are different. In the soft X-ray range the flux from lv  shocked nebulae dominates
over the WD bb only for \Vs$\geq$ 150 \kms. 

 The comparison of the fluxes calculated at the nebula  with the continuum SED observed 
at Earth indicates that
hv shocks   have a radius smaller than that of  lv  ones (Table 2, last row), so
the fluxes observed at Earth of  hv nebulae  would be lower   by  several orders of 
magnitude than those  corresponding to
lv  nebulae and  WD bb radiation. However, both the latter fluxes  are absorbed by the envelop of debris
accompanying  the explosion.  
The drop in the X-ray counts at the outbursts can thus  be explained by the different origin of the
X-ray.

Summarizing, at outbursts, the X-rays are  emitted from the shock accompanying the 
blast wave (which  absorbs the WD bb flux). The observed flux 
is relatively low because it is diluted by  the small $\eta$ factor.
 In the quiescent and transition phases
the soft X-ray data (observed at similar phases but at different epochs by Skopal et al) are  
reproduced by the WD  bb  with  a contribution  from  the  bremsstrahlung  
which dominates for shock velocities  $\geq$ 150 \kms.
The optical-UV flux  is bremsstrahlung emitted from downstream nebulae  
photoionized by the WD bb flux.

\begin{figure}
\begin{center}
\includegraphics[width=0.50\textwidth]{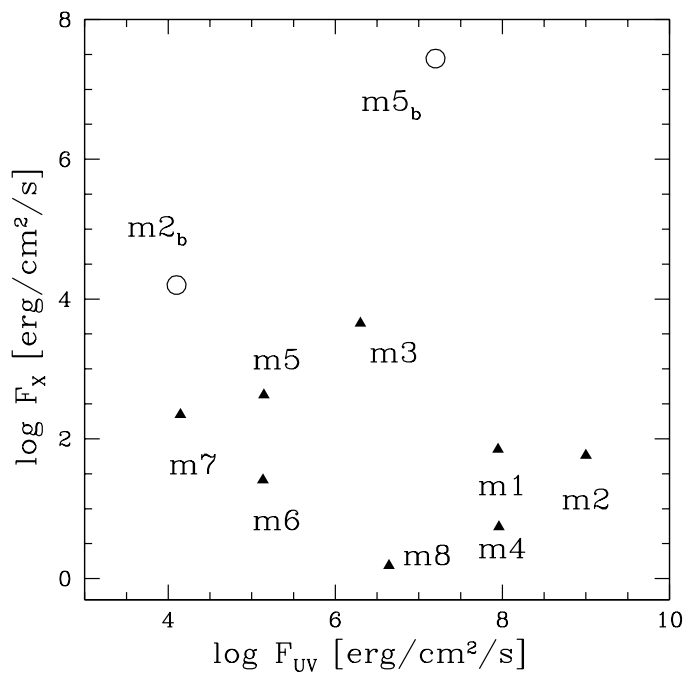}
\caption{ F$_X$ versus  F$_{UB}$
calculated at the nebula by models which explain the line and continuum spectra.
Filled triangles: low-velocity models;  open  circles:  high-velocity models.
}
%label{fig:ions}}
\end{center}
\end{figure}

\subsection{Relative abundances}

The abundances of the heavy elements relative to H which
explain the observed line ratios are given in Table 2 (rows 6-9).
We could model only the elements corresponding to observed lines.

The results show  a strong  under-abundance of the heavy elements
relative to the solar ones  (C/H=3.3 10$^{-4}$, N/H=9.1 10$^{-5}$, O/H=6.6 10$^{-4}$, 
Mg/H=2.6 10$^{-5}$, Nussbaumer et al. 1988).

 Nussbaumer et al. (1988) compared the  O/N  and C/N  ratios  for different objects, e.g.
symbiotic stars, planetary nebulae, novae, etc.
Our results show that AG Dra has  C/N  comparable with other symbiotic stars, but higher O/N.
Moreover  O/N ratios  are even  outside the ranges covered by the other objects.

We find C/O varying from 0.05 to 0.13, while the solar abundance ratio
is  0.5.  The N/H ratio is underpredicted
by a factor of 4 relative to  the solar, while C/H is  lower than solar
by a factor $>$ 10. Also O/H is lower than the solar by a factor
of $\sim$2.7 in average, while Mg/H  varies between solar and 0.1  solar depending on the phase.
 Mg can be trapped into  silicate grains, so the
 different abundance ratios indicate an inhomogeneous distribution of dust.
 We cannot comment about the Si/H relative abundance, because no Si lines are given in the spectra.
SiIV] line is blended with OIV] in the 1400 \AA ~  multiplet. The calculated spectra show that
 the intensity of the SiIV]
line is always lower than that of  OIV] or   similar in the case of  model m4.
For m4  N/H is rather solar,
while C and O are depleted strengthening  the hypothesis that C and O are trapped into  CO molecules.

The origin of the  high depletion of C, N, and O relatively to H and He  is most probably
due to  inclusion  in dust grains and/or molecules (Angeloni et al. 2007a).
Dust grains are coupled  to the gas by the magnetic field  downstream of  shock-fronts. They   
are heated by  radiation from the WD and collisionally by the gas. Starting  by an average
 grain radius a$_{gr}$=0.2 \mum,
Fig. 2 (top panels) shows that the  grains reach a temperature  $<$ 2000 K and survive  
sputtering and evaporation (which are calculated consistently with the temperature of the gas)
in the transition and quiescence phases,  but are completely destroyed
downstream of  high velocity shocks.  

Fig 3 shows that  dust reprocessed radiation
is  hidden  throughout the SED  by the bb flux from the RG. 
Adopting $d/g$ ratios (Table 2) similar to those of the
ISM (4.1 10$^{-4}$ by mass which corresponds to 10$^{-14}$ by number)
 dust emission does not contribute to the continuum sensibly.
 At the dates 6/12/1994 and 28/7/1995 a $d/g$  $\sim$10$^{-15}$ has been adopted.
 However,  these results are based on  the ISO observations
at  date 29/July/1996. Actually, dust emission can  vary with phase,
as was found for e.g. RS Ophiuci by Rushton et al. (2010).

Oxygen is depleted because it is trapped in unsputtered silicate grains and in CO molecules, 
while carbon is much more depleted (by a factor of 3) because it is trapped into graphite grains, PAHs, and 
diatomic molecules like CO, CN, CS, etc. (Contini \& Shaviv 1982).
Formation of dust species downstream of shocks  is  characteristic of the RG.
Actually,   {\it shock-excited} far-IR emission of CO
was  detected by Reach \& Rho (1998)
with  the ISO Long-Wavelength Spectrometer  from the  SN remnant 3C 391.

As a result, the depletion of C, N, O 
 and the $\sim$1 year fluctuation period  observed  during the
high outbursts in the U band and  in the V band 
strengthens the hypothesis that this short  period depends  
on  the sudden  disturbance of the dynamical field between the stars
by the   collision of the dusty shells ejected from the RG with the wind from the WD.

\section{Discussion and concluding remarks}

In  previous sections we  have demonstrated
that the bremsstrahlung emitted downstream of
shocked nebulae  in  AG Dra SS  contributes to  the soft X-rays  at different phases.

GVIG99 report that 
after the quiescence period from 1969 to 1980 and  the active phase
from November 1980 and beginning 1983, a new minor outburst took place in March 1985
followed by another one  after 340 days in January 1986.
After 1986 a 8 years period of quiescence followed until the major active phase started
in June 1994 with at least five   peaks
in June 1994, June 1995, July 1996, and June 1997, and one in summer
 1998. They are  separated by nearly the same time interval of about 360 days.
Another small light maximum is present 330 days before  the first maximum (Fig. 1, top).

We have  reproduced the UV-optical spectra at  different epochs by model calculations (Tables 1 and 2):
Nov 4 1985 (m1), Jan 18 1986 (m2), Apr 30 1986 (m3), Feb 27 1993 (m4), Jul 4 1994 (m5), 
Dec 6 1994 (m6), Jul 28 1995 (m7), and Feb 14 1986 (m8). 
The bremsstrahlung calculated by model m1 refers to a transient phase, m2 to a maximum,
m3 to the end of the maximum, near quiescence,
m4 to quiescence, m5 is close to maximum, m6 corresponds to a transient phase in terms of a small
depression between two maxima, m7 is close to
maximum, and m8  represents a  quiescence phase.

The classification in quiescence, maximum,  and transient phases by Skopal et al,  provides only a
schematic picture of the physical conditions. We believe that the conditions change
from burst to burst, from quiescence to quiescence, etc., because 
collision of the fast wind from the WD with the RG wind  between the stars and 
with the medium inhomogeneities 
on the side opposite the RG, at outbursts, leads
to a network of shock-fronts which  varies  from time to time.

Accepting that  Skopal et al.  (2009, fig. 2) continuum data  can be applied at  different epochs,
Fig. 3 shows that  the soft X-rays would
correspond to the bb flux from the WD in transient and quiescence epochs
for shocks  with  \Vs $<$ 150 \kms.
During the 18/1/1986 burst the X-rays originated
from the high velocity shock and  could be observed also  beyond 1 keV.
After the burst,
on 30/4/1986, the soft X-rays corresponding to the bb flux are comparable
with the bremsstrahlung downstream of the reverse shock  with \Vs=150 \kms.

During  the 4/7/1994
and 28/7/1995 bursts  the bremsstrahlung  downstream of the  reverse shock corresponds to
 the soft X-ray range, while  downstream of the blast wave shock it  covers also   the X-ray range
($>$ 1 keV).
There are no contemporary data which could confirm our prediction in the X-ray range.

The  fluctuations in the U band  during the active phases  between 1995 and 1999
show a $\sim$360 days periodicity (Fig. 1, top diagram)  which is
 similar to  the red giant pulsation  period ($\sim$1 year). Pulsations
provoke  dusty shells ejection.
Collisions of the shells with the WD wind   give origin to the   sequence of peaks  during the
active phases. This is  connected   to  the   gas and dust composition
in shocked nebulae. Calculation results  show
a depletion of C by a factor $>$ 10, a depletion of O by a factor of 2.7 and of N
by a factor
of $\sim$4, while Mg is more variable. The depleted elements are
trapped into dust grains (silicates and graphite) and/or into diatomic  molecules
(CO, CN, CS, etc.).

The  prerogative of  observing the effect of the RG pulses throughout the light curve  in U
belongs mainly to systems which are seen nearly face-on, like AG Dra. Most information
is lost in edge-on systems due to eclipses not only by the two main stars, but also
by the different nebulae and shells (e.g. CH Cyg, Contini et al. 2009c) within the system.

Fig. 3 diagrams show that the continuum in the optical-UV  range is emitted by the shocked nebulae
between the stars. 
The low-frequency tail of the WD bb flux is seen  at  $\nu$ $>$ 10$^{15}$ Hz.

It was concluded by Skopal et al. (2009) that the super soft X-ray bump
represents directly the WD bb flux. 
The peak of the WD bb between $\sim$3 10$^{15}$ and
3 10$^{16}$ Hz  is not reproduced by shock dominated models. Actually, it was
neither confirmed by the observations due to the strong absorption in this range.
Moreover, the H  column density N$_{H}$ $\sim$3 10$^{20}$ cm$^{-2}$
adopted by Skopal et al. (2009) to correct the data to  fit the Planck function,
can sensibly vary with time in  a complex system such as a symbiotic one, if 
 column densities from the emitting nebulae  are accounted for.

The anti-correlation between the
X-ray and  UV fluxes, particularly at outbursts, was explained by the obscuration of the WD
by the envelop of
matter ejected at outburst which absorbs the soft X-ray emission (Skopal et al. 2009).
We suggest that the   absorbing envelop 
corresponds to the  fragmented  nebula downstream of the high velocity shock accompanying the blast wave,
which was successfully invoked to explain
some SS features, such as the broad Ly$\alpha$ line in CH Cyg (Contini et al. 2009a).
 The fast debris  emit  the broad line spectra (Sect. 3.2) and the X-ray bremsstrahlung
observed at  outbursts.

Finally, the results presented for AG Dra are in agreement with
Smith et al. (1996) who claim by simple arguments based on the luminosity function of solar and 
low-metallicity K giants that yellow symbiotic stars arise among low-metallicity stars, because 
low-metallicity K giants have a larger mass loss rate,  thus triggering symbiotic-like activity
in a binary system.
Generally yellow SS show  IR emission from dust (e.g. HD 330036, Angeloni et al. 2007b) which cannot be 
observed in AG Dra because it is hidden the bb RG flux throughout the SED.
However, we have found that  dust is present between the stars  with a  non-negligible dust-to-gas ratio 
($\leq$ 4 10$^{-4}$ by mass).

\section*{Acknowledgments}
We are  grateful to  the anonymous referee for his  comments which  improved the presentation of the 
paper. We thank Dina Prialnik  and Elia Leibowitz for  interesting  conversations
and  Sharon Sadeh for helpful advise. RA acknowledges a grant from the FONDECYT Project N. 3100029.

%\bibliographystyle{elsarticle-num}

\end{document}